\documentclass[10pt]{iopart}   % 10pt for double column and 12 for single

\usepackage{graphicx}%
\usepackage{epstopdf}
\usepackage{color}
\usepackage{upgreek}
\usepackage{tablefootnote}

\begin{document}

\title[]{Full cycle, self-consistent, two-dimensional analysis of a packed bed DBD reactor for plasma-assisted $\mathrm{CO_{2}}$ splitting: spatiotemporal inhomogeneous, glow to streamer to surface discharge transitions.}
%{Two dimensional analysis of a packed bed DBD reactor for plasma-assisted $\mathrm{CO_{2}}$ splitting: discharge initiation and full-cycle spatiotemporal characteristics of streamers and MDs}

\author{K Kourtzanidis}

\address{Chemical Process \& Energy Resources Institute (CPERI), Centre for Research \& Technology Hellas (CERTH),
6th km Charilaou-Thermi, Thermi, 57001 Thessaloniki, Greece}
\ead{kourtzanidis@certh.gr}

%\begin{indented}
%\item[]March 2018
%\end{indented}
%\begin{indented}
%\item[]March 2018
%\end{indented}

\begin{abstract}
We investigate the full-cycle operation of a coaxial Packed Bed Dielectric Barrier Discharge (PB-DBD) reactor operating in pure $\mathrm{CO_{2}}$. The reactor is packed with high permittivity dielectric rods and is analyzed with a two-dimensional (2D) self-consistent plasma model. We show that the PB-DBD operation is governed by both glow and volume/surface streamer discharges, forming alternatively and non-uniformly inside the gas volume. The presence and surface charging of the dielectric rods and dielectric layer is crucial for the initiation, propagation, annihillation and afterglow of these microdischarges. Our calculations show maximum electron and  $\mathrm{CO_{2}^{+}}$ densities in the order $10^{20}$ $\mathrm{m^{-3}}$, an average discharge power of 353.42 W/m, microdischarge peak currents in the order of 50-400 A/m, total half-cycle plasma charge of around 6 $\mathrm{\mu}$C/m. Dominant negative ions are found to be  $\mathrm{CO_{3}^{-}}$. CO molecules and O atoms are mainly formed during the MDs development and the streamer-surface ionization waves. Molecular oxygen ($\mathrm{O_{2}}$) is preferentially formed during the glow, current-decaying and afterglow phase of each microdischarge. The spatially average reduced electric field inside the reactor lies in the 20-100 Td range. Each MD, presents distinct non-uniform and non-repeatable glow and volume/streamer discharges owed to the non-uniform surface charging processes which dictate the complex spatial distribution of produced neutral (and ion) species. These detailed results shed light on crucial, largely non-uniform plasma spatiotemporal characteristics that can help design efficient PB-DBD reactors for $\mathrm{CO_{2}}$ splitting and beyond, while emphasizing the important insights obtained by 2D simulations which can not be captured with 0D-global or 1D models.

\end{abstract}

%\submitto{\PSST}
\ioptwocol

\maketitle

\section{Introduction}\label{sec:intro}

In the pursue of technologies and green processes that can substantially reduce Green House Gas (GHG) emissions, plasma-assisted $\mathrm{CO_{2}}$ splitting has been proposed as a promising alternative to more conventional, thermodynamically-driven~\cite{nigara1986production} and/or electrochemical methods~\cite{xie2019electrochemical}. Indeed, several authors have demonstrated and summarized the potential of plasma discharges as an interesting approach inside the Carbon Capture Utilisation and Storage (CCUS) value-chain~\cite{centi2021plasma, bogaerts2020plasma, marcantonio2022non}: captured $\mathrm{CO_{2}}$ from the atmosphere or energy intensive industries can be directly converted into $\mathrm{CO}$ and $\mathrm{O_{2}}$ via direct electron impact dissociation and a ladder-climbing dissociation process through vibrational excitation and vibrational-vibrational (VV) collisions. The aforementioned products are not considered GHGs and as such, climate change and global warming could be mitigated by substantially reducing $\mathrm{CO_{2}}$ content in the atmosphere. In addition, these products lay the path for synthesizing added-value fuels and chemicals such as biogas, methanol, propane and jet fuel, contributing to the Utilisation (and valorisation) side of CCUS. 

To this end, a series of plasma sources and reactors have been proposed for  $\mathrm{CO_{2}}$ splitting: corona discharges, dielectric barrier discharges (DBD), microwave plasma and thermal plasmas display significant advantages as well as drawbacks regarding conversion and overall power efficiency,  depending on the operating conditions and reactor configurations. Focusing on cold plasma discharges, DBDs~\cite{paulussen2010conversion} have been recognised as a less energy efficient plasma source under low pressure conditions due to the promotion of direct electron impact dissociation rather than the ladder-climbing dissociation pathway. The latter is considered as the most energy efficient pathway for $\mathrm{CO_{2}}$ splitting, requiring the minimum amount of energy for bond-breaking, i.e. 5.5 eV (in contrast to the minimum of 7 eV threshold for the direct dissociation pathway). Still, under atmospheric conditions, DBDs offer the simplest, more robust and highly modular design compared to other plasma sources while they can reach conversion rates higher than 30\%~\cite{aerts2015carbon}. The ease of connection of DBD reactors in series or in parallel operation is yet another advantage of such very well-studied plasma sources in ambient conditions, towards increased conversion efficiency~\cite{vertongen2022enhancing}.  In addition, DBDs can be easily arranged in a co-axial configuration and packed with catalysts towards packed bed DBD (PB-DBD) reactors\cite{zhu2022co2, rad2023enhancement}, leveraging plasma-catalyst synergies. Focusing on the latter, plasma catalysis remains an emerging field with large unexplored potential. Coupling plasma reactors with catalysts present significant synergetic effects in multiple scales, that can further boost conversion efficiency~\cite{bogaerts20202020, carreon2019plasma}. On one hand, (typically dielectric) catalyst supports and active materials directly impact the formation and type of plasma discharges encountered in packed bed reactors through electric field enhancement/alteration, dielectric charging etc. On the other hand, plasma itself seems to substantially alter the catalyst activation and performance through changes in oxidation states, work function, activation barrier and overall morphological chemical and surface properties.\cite{neyts2015plasma} Concerning PB-DBDs, the influence on $\mathrm{CO_{2}}$ conversion of packing and catalyst materials, gap and sphere size combination as well as overall reactor setup and flow rates has been extensively studied~\cite{uytdenhouwen2018packed, michielsen2017co2, mei2017atmospheric}. In Ref~.\cite{rad2023enhancement}, an increase of $\mathrm{CO_{2}}$ splitting and energy yield has been demonstrated using glass supports with $\mathrm{CeO_{2}}$ coating. The sensitivity of synthesis induced differences in the morphology of millimeter-sized SiO2@TiO2 packing spheres on a PB-DBD $\mathrm{CO_{2}}$ conversion rate has been shown in Ref.~\cite{kaliyappan2021probing} while different core-shell materials for tuning of conversion efficiency are explored in Ref.~\cite{uytdenhouwen2020potential}. Increased conversion with $\mathrm{\gamma-Al{2}O_{3}}$ supported potassium intercalated KuCN/AO catalysts (compared to a pure DBD/$\mathrm{\gamma-Al{2}O_{3}}$ system) has been demonstrated in Ref.~\cite{lu2021co2}. The influence of operating conditions was investigated in Ref.~\cite{xu2018co2} for a barium titanate packed-bed plasma reactor operating in pure $\mathrm{CO_{2}}$ and $\mathrm{CO_{2}}$/Ar mixtures. The above studies are surely a non-exhaustive list of the expanding literature on plasma assisted $\mathrm{CO_{2}}$ conversion but demonstrate the increasing interest of the technology as well as the complexities of plasma-catalyst synergies.

Numerically, the elucidation of $\mathrm{CO_{2}}$ conversion and various reaction pathways in atmospheric pressure (AP) PB-DBD reactors has been mainly based on 0D-global modeling efforts~\cite{bogaerts2017plasma, aerts2012influence, kozak2014splitting}. Nevertheless, the presence of dielectric structures (catalyst supports, dielectric layers), specific reactor designs and multi-filament nature of AP-DBDs, limit the applicability of 0D modeling approaches due to their inherent assumptions~\cite{alves2018foundations}. PB-DBDs in ambient and moderately high pressure conditions are governed by multiple discharge regimes ranging from glow and streamers which transit to MicroDischarges (MD) as well as surface ionization waves (SIW)~\cite{chirokov2005atmospheric, brandenburg2017dielectric}, often exhibiting complex self-organized patterns~\cite{boeuf2012generation}. The initiation and development of each phase are dictated by spatiotemporal parameters linked to the applied voltage waveform (magnitude, frequency) as well as the packing arrangement, materials and reactor design. 0D-models typically consider periodically repeatable and identical MDs in the same spatial regions of the reactor. Incorporating into such models the MDs frequency, dimensions and gas flow rate (i.e. residence time)~\cite{snoeckx2013plasma, snoeckx2017quest, alliati2018plasma} allows for an extrapolation of the obtained results in multiple AC cycle timescales but the assumptions made on the MDs nature render these results uncertain especially for PB-DBDs operating in the filamentary regime. An excellent Monte-Carlo based study demonstrating the complexity of PB-DBDs physics (including gas flow phenomena) as well as the sensitivity of common modeling assumptions can be found in Ref.~\cite{van2021spatially}. The validity of 0D plasma kinetics models would be greatly enhanced if the necessary input parameters would be based on accurately predicted, non-uniform and time-dependent in nature plasma densities, discharge power and/or spatiotemporal evolution of generated species. Moreover, higher dimensional simulations can help understand fundemantal plasma processes in PB-DBDs, eventually related to conversion efficiency. 

 To the best of our knowledge, the only higher-than-zero dimensional numerical work on AP-PB reactors in $\mathrm{CO_{2}}$ is the one from Ref.~\cite{van2017bead}. Therein, the authors have studied a particular, Helium filled, PB-reactor geometry, effectively representing a planar, circular DBD packed with spherical and torrus-like beads, revealing interesting results on multi-discharge phenomena inside a relatively high-frequency (23.5 kHz) AC cycle. Despite several interesting two-dimensional (2D) numerical studies of plasma-dielectric interactions and DBDs/PB-DBDs under ambient conditions and various gas environments (e.g., see Ref.~\cite{kruszelnicki2020interactions, kruszelnicki2016propagation} for humid air, Ref.~\cite{ran2022numerical} for helium), literature lacks a detailed 2D full-cycle analysis of a PB-DBD reactor operating in ambient $\mathrm{CO_{2}}$, let alone a study in low AC frequencies (1-10 kHz). Such analysis is extremely demanding in terms of computational resource allocation.  Advanced numerical algorithms are necessary to capture efficiencly and accurately, the highly multi-scale nature of the problem, in a reasonable computational time.

To this respect, this work presents two-dimensional (2D) self consistent plasma simulations of a coaxial packed bed DBD reactor operating in pure $\mathrm{CO_{2}}$. The filamentary nature of the discharge in a loosely PB-DBD and the influence of the streamers in the conversion efficiency has been recently demonstrated experimentally in Ref.~\cite{zhu2022co2}. In this reference, the authors show that a DBD reactor packed with simple, high-permittivity, dielectric rods (no active material) can significantly enhance the $\mathrm{CO_{2}}$ conversion and efficiency. The authors attribute this enhancement to a so-called stable streamer formation between the inner electrode and dielectric rods based on ICCD discharge imaging, while emphasizing the importance of streamer intensity and length on $\mathrm{CO_{2}}$ splitting efficiency. The existence of multiple MDs (evident by the charge measurements of the aforementioned article) inside an AC period and spatial inhomogeinity of the plasma formation might play an important role which can not be easily captured by typical, experimentally used, measurement techniques. To this respect, the focus of this work is on the multiple discharge regimes encountered inside a full AC cycle (2 kHz) and the spatiotemporal behaviour of such discharges which will eventually dictate the splitting distribution and overall efficiency over multiple AC cycles. The structure of the article is as follows. First, we present briefly essential modelling aspects and the simulation conditions (Sec. \ref{sec:modeling}). Then numerical results of the plasma initiation, development and streamer/MD formation in both half-cycles are presented and compared to experiments (Sec. \ref{sec:results}). \color{black}In Sec.~\ref{sec:discussion}, we elaborate on the main assumptions of the physical model used in this study and the expected influence of those, on the numerical results. \color{black} Finally in Sec. \ref{sec:conclusion}, we summarize our findings and provide future directions.

\section{Modeling and computational aspects}\label{sec:modeling}
A fluid (continuum) description of the plasma discharge is used. The main aspects of the physical model and numerical solver details have been described elsewhere (COPAIER solver~\cite{kourtzanidis2020self, dufour2015numerical} - a multi-species, multi-temperature plasma fluid solver) while the solver has been recently expanded with additional capabilities. In brief, Poisson equation is self-consistently coupled to a drift-diffusion-reaction equation derived for each species continuity.  The species number flux is given by the momentum balance equation which is approximated by the drift-diffusion equation.  In this work, the reaction and transport rates depend on the reduced electric field under the commonly used (valid for collisional, high-pressure conditions) Local Field Approximation. The density of the background species is given by the ideal gas law. Surface charge accumulation on dielectric surfaces is taken into account by assuming that particles stick to the dielectric surface and no diffusion occurs. Secondary electron emission (SEE) from electrode/dielectric surfaces is included. A ballast resistor incorporated into a simple electrical circuit model is used to mimic the High Voltage (HV) power supply. Photoionization is not included in the current case as its relevance in pure CO2 discharges under typical glow/streamer conditions (mean electron energies in the 1-10 eV range) is assumed negligible\cite{bagheri2020simulation, levko2017particle} (see also Sec.~\ref{sec:discussion}). The numerical solver uses adaptable time-stepping,  implicit time-integration~\cite{besse2023implicit} for the diffusion-advection-reaction system obeying the dielectric relaxation time scales, efficient finite element (for Poisson equation) and finite volume (for continuity equations) algorithms and MPI parallelism. The Scharfetter-Gummel flux scheme~\cite{scharfetter1969large} is used here to discretize the convection-diffusion operator of species continuity equations. The chosen numerical schemes provide a good compromise between CPU efficiency and accuracy for the purpose of this study (note that COPAIER~\cite{kourtzanidis2020self} allows the user to choose different schemes and physical models, providing flexibility in physical, spatial and temporal order of accuracy and consequently simulation cost).

\begin{figure}[h]
\centering
\includegraphics[width=1\linewidth]{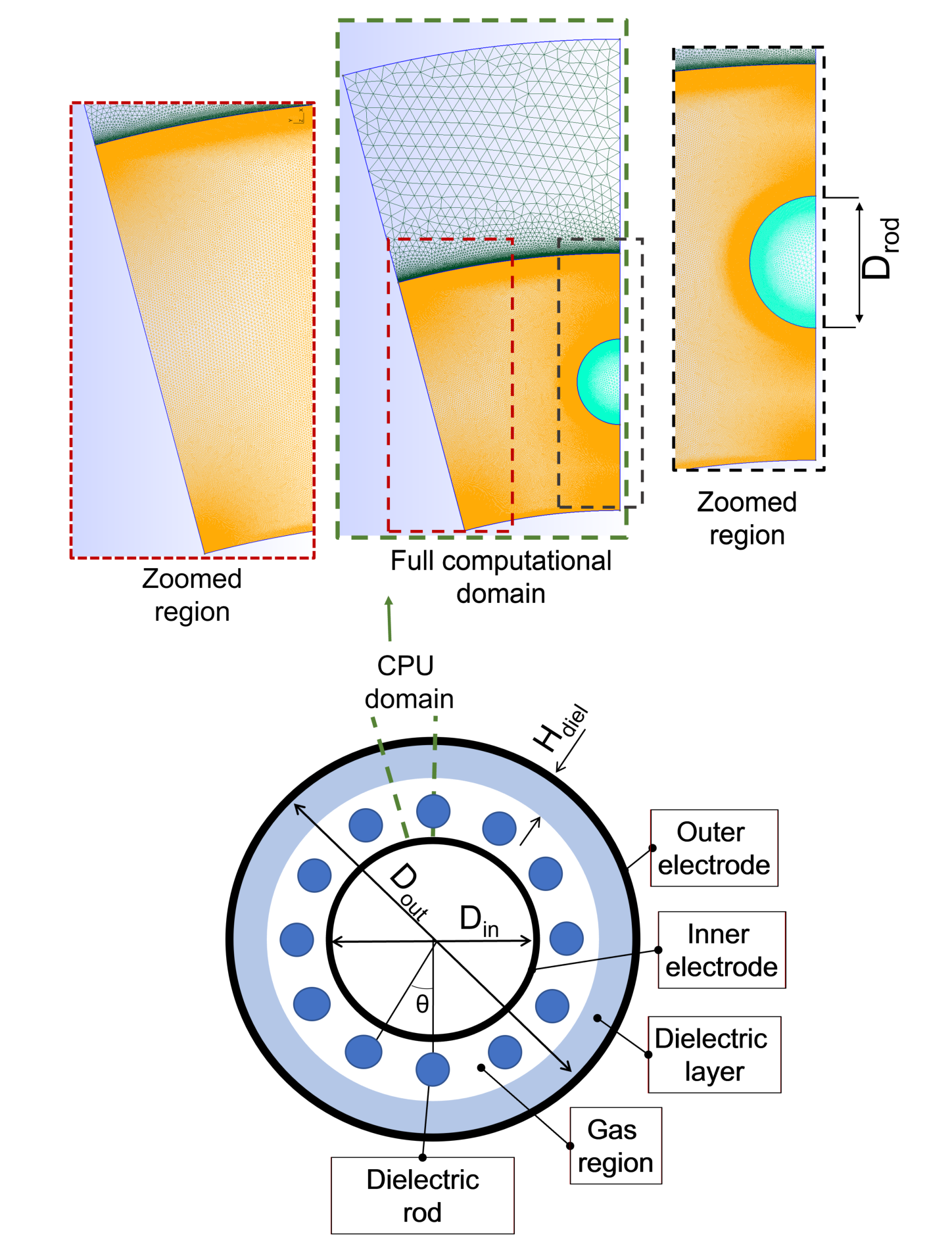} 
\caption{\label{fig:domain} \color{black} Bottom: The full two-dimensional domain consisting of 12 rods (dimensions not to scale). The angle phi controls the number of rod packing. The computational domain is limited by the green dashed lines with symmetric (periodic) boundary conditions. Top: Computational mesh (middle) and zoom in the region of the dielectric rods (right - black dashed lines) and the region between the rods (left - red dashed lines). Note the refinement regions around the rods and near the dielectric and inner electrode regions. \color{black}
}
\end{figure}

     The simulated geometry corresponds to the experimental study of Ref.~\cite{zhu2022co2} and the 12 dielectric rods case. It consists of a coaxial DBD reactor, with a ground rod electrode at its center and a high-voltage electrode wrapped around a quartz tube with dielectric permittivity of 4.5. The radius of the ground inner electrode is $R_{in}=7$ mm. The radius of the outer electrode is $R_{out}=12.5$ mm.  The thickness of the dielectric barrier is $H_{diel}=2.5$ mm so that the discharge is confined in a 3 mm gap. The coaxial DBD is packed with twelve (12)  $\mathrm{ZrO_{2}}$ rods with a dielectric permittivity of 25, centered in the middle of the gap and equally spaced. The diameter of the rods is 1 mm. The radius of the dielectric rods centers is thus 8.5 mm and the angle between each rod centers is $\theta=60^{\circ}$. In the 2D approximation of the experimental domain, the axial length of the reactor which is 35 mm (note a possible error in Ref.~\cite{zhu2022co2} where it is noted that after the holders' positioning, the discharge length is reduced by half, so from 70 mm it should go down to 35 mm and not 3.5 mm as mentioned in the manuscript)  is only used to scale the discharge current. Fig.~\ref{fig:domain} illustrates a 2D cut plane of the reactor's domain, the reduced simulation domain where symmetric boundary conditions have been used on the top and bottom boundaries, as well as the computational mesh in different zones of the domain.  The computational mesh is an unstructured one created in the Gmsh~\cite{geuzaine2009gmsh} software, properly refined near the electrode and dielectric layer edges as well as near the dielectric rods.  The total mesh consists of 352.947 triangular elements. The minimum cell size is 2.5$\mathrm{\mu}$m at the `boundary layer' zones at the dielectric layer surface and at the poles of the dielectric rods. We emphasize the importance of a locally refined mesh in order to capture accurately the streamer propagation, cathode layers formation and relaxation, SIWs and full MD development in different regions of the reactor. The domain has been decomposed in 22 sub-domains and run on a multi-proc CPU machine (AMD EPYC 7352). \color{black}The computational time for one full, 2 kHz, AC cycle was approximately 40 days.\color{black}

The operating AC frequency is 2 kHz with an applied voltage of 13.5 $\mathrm{kV_{pp}}$. In Table~\ref{tabtwo}, we summarize the geometrical and operational conditions for the simulations (see also Fig.~\ref{fig:domain}).
 
\begin{table}
\caption{\label{tabtwo}Geometrical and operational conditions for the simulations. See also Fig.\ref{fig:domain} for reference on dimensions.} 
\begin{indented}
\lineup
\item[]\begin{tabular}{@{}*{2}{l}}
\br                              
$\0\0\textbf{Frequency}$&f=2 kHz\cr
%$\textrm{Mesh support}$&$\textrm{Accuracy}$&\m$\textrm{LFA/LEA}$&\m$\textrm{Photoionization}$&$\textrm{\# of species and reactions}$&$\0\textrm{Parallelization}$\cr 
\mr
$\0\0\textbf{Voltage}$&$V_{max}$ = 13.5 kV\cr
\mr
$\0\0\textbf{Inner electrode diameter}$&$D_{in}$=$14$  mm\cr
\mr
$\0\0\textbf{Outer electrode diameter}$&$D_{out}$ = $25$ mm\cr
\mr
$\0\0\textbf{Dielectric layer thickness}$&$H_{diel}$ = $2.5$ mm\cr
\mr
$\0\0\textbf{Dielectric rods diameter}$&$D_{rod}$ = $1$ mm\cr
\mr
$\0\0\textbf{Angle between rods}$&$\theta$ = $30^{\circ}$ (12 rods)\cr
\mr
$\0\0\textbf{Rel. permitivity of dielectric layer}$& $\epsilon_{r, diel}$=$4.5$\cr
\mr
$\0\0\textbf{Rel. permitivity of dielectric rods}$& $\epsilon_{r, rod}$=$25$\cr
\mr
$\0\0\textbf{Electric circuit resistance}$&100 $\Omega$\cr
%\0\02D/2.5D&Structured/Unstructured/Hybrid  &2nd order (space and time)&Both&Yes &\0Arbitrary (tabulated)&\0Yes (MPI)\cr
\br
\end{tabular}
\end{indented}
\end{table}

\subsection{Plasma chemistry model}\label{sec:chemistry}
We choose to use a reduced $\mathrm{CO_{2}}$-plasma chemistry, where nine (9) modeled species are included, consisting of electrons, positive  $\mathrm{CO_{2}^{+}}$ ions, negative  $\mathrm{O^{-}}$, $\mathrm{O_{2}^{-}}$ and $CO_{3}^{-}$ ions, neutral $\mathrm{CO}$,  $\mathrm{O_{2}}$ and  $\mathrm{CO_{2}}$ molecules as well as atomic oxygen $\mathrm{O}$. Fourteen (14) reactions are taken into account which are listed in Table~\ref{tabfour}: direct electron impact ionization of $\mathrm{CO_{2}}$ (R1). Electron impact dissociation of $\mathrm{CO_{2}}$ (R2) and $\mathrm{O_{2}}$ (R5). Dissociative attachment to $\mathrm{CO_{2}}$ (R3) and $\mathrm{O_{2}}$ (R6). Electron-ion recombination for positive $\mathrm{CO_{2}^{+}}$ ions (R4). Two seperate three-body electron attachment reactions forming $\mathrm{O_{2}^{-}}$ and $\mathrm{O^{-}}$ ions (R7-R8).  Ion-neutral three-body transformation reaction of negative $\mathrm{O^{-}}$ ions with $\mathrm{CO_{2}}$ forming $\mathrm{CO_{3}^{-}}$ (R9). Ion-ion recombination of positive $\mathrm{CO_{2}^{+}}$ and negative $\mathrm{CO_{3}^{-}}$ ions  (R10) as well as negative $\mathrm{O_{2}^{-}}$ ions  (R11) . Three-body neutral recombination of O and CO atoms to form $\mathrm{CO_{2}}$ (R12) as well as O atoms to form $\mathrm{O_{2}}$ (R13). Electron detachment from $\mathrm{CO_{3}^{-}}$ ion in collisions with neutral $\mathrm{CO}$ molecules (R14). 
%The ion-ion recombination reaction (R7) assumes that  $\mathrm{O^{-}}$ ions rapidly (instantaneously) recombine (three-body reaction with $\mathrm{CO_{2}}$) to  $\mathrm{CO_{3}^{-}}$ ions which then recombine with positive $\mathrm{CO_{2}^{+}}$  ones. 

Detachment reactions involving $\mathrm{O^{-}}$ ions (with $\mathrm{CO}$ and $\mathrm{O_{2}}$), are neglected. The production of ozone ($\mathrm{O_{3}}$) is also neglected in this model. Three-body reactions (R7-R8, R12-R13) involve only $\mathrm{CO_{2}}$ as third species (M). The assumptions made in the reduced chemistry model can be justified due the low dissociation degree expected in the first cycles of the atmospheric pressure DBD operation, which is the main focus of this work. The chemistry set is similar (while simpler) to the benchmark one from Ref.~\cite{wang2018modelling}, while it includes $CO_{3}^{-}$ ions. We argue that the inclusion of these ions might be important for the charge balance of the discharge as R9 will dominate under the high-pressure conditions considered here and thus $O^{-}$ ions will quickly be transformed to $\mathrm{CO_{3}^{-}}$.  A detailed chemistry and analysis of reaction mechanisms fall outside the scope of this work. 
 
Transport and reaction rate coefficients for R1-R3 and R6-R7 have been calculated using the two-term Boltzmann solver, BOLSIG+~\cite{hagelaar2005solving} including electron-electron, electron-ion and superelastic collisions and under the density gradient expansion, spatial growth (SST) model. We assume a gas composition of 92.19\% $\mathrm{CO_{2}}$, 7.8\% of $\mathrm{CO_{2}}$ (0 1 0) vibrational state (see below) and 0.1\% $\mathrm{O_{2}}$ (based on relevant populations found in DBDs using global modeling - see e.g. Ref.~\cite{aerts2015carbon}). The ionization degree is assumed $10^{-6}$, the electron density $10^{19}$ $\mathrm{m^{-3}}$ and the gas temperature 300 K. Cross-sections for $\mathrm{CO_{2}}$ used as inputs in BOLSIG+ have been taken from the IST-Lisbon database~\cite{lisbon} which is based on the work of Ref.~\cite{grofulovic2016electron} and includes 17 electron-neutral scattering cross sections describing dissociative attachment, effective momentum transfer, ionization, vibrational excitation and superelastic de-excitation of the $\mathrm{CO_{2}}$ (0 1 0) vibrational state. The latter (0 1 0) state has a relative population of about 8\% even at room temperature~\cite{grofulovic2016electron}.
state. Electron impact dissociation is assumed to be linked with the 7 eV vibrational level, as proposed in Ref.~\cite{bogaerts2016modeling}. We note here the on-going debate about the best cross-sections for electron-impact dissociation and redirect the readers to Ref.~\cite{polak1976electron, pietanza2016electron, leclair19941s, grofulovic2016electron} for further reading. We choose not to track individual vibrational levels in the simulations, and so we neglect ladder-like dissociation processes. In typical DBD conditions, it has been proven that direct electron impact dissociation is the main dissociation mechanism~\cite{kozak2014splitting}, which contributes to more than 90\% compared to other pathways. For $\mathrm{O_{2}}$, we have used the cross-sections from the Phelps database~\cite{phelps} which includes 17 cross-sections describing two-body and three-body attachment, effective momentum transfer, rotational vibrational and electronic excitation as well as ionization. We assume that electron impact dissociation proceeds mainly via electronic transition to the Herzberg states. Ion mobilities for $\mathrm{CO_{2}^{+}}$ and $\mathrm{CO_{3}^{-}}$ (in $\mathrm{CO_{2}}$) have been taken from LxCAT swarm data (Viehland database~\cite{viehland1995relating}), for $\mathrm{O_{2}^{-}}$ (in $\mathrm{O_{2}}$) from the same database, while for $\mathrm{O^{-}}$ from Ref.~\cite{harrison1972ion, okuyama2012measurement}. Constant diffusion coefficient of $\mathrm{1.9 \times  10^{-5}}$ $\mathrm{m^{2}/s}$, $\mathrm{1.4 \times  10^{-5}}$ $\mathrm{m^{2}/s}$, $\mathrm{1.9 \times  10^{-5}}$ $\mathrm{m^{2}/s}$ and $\mathrm{2.88 \times  10^{-5}}$ $\mathrm{m^{2}/s}$ has been used for CO, $\mathrm{CO_{2}}$, $\mathrm{O_{2}}$ (assuming air as background gas) and $\mathrm{O}$ (calculated for atmospheric conditions from Ref.~\cite{cenian1995modeling}) neutral species, respectively.

 Pure atmospheric pressure CO2 is considered as background gas, neglecting gas flow dynamics as the timescales of gas convection (in the perpendicular direction of the computational domain) are much larger than the AC cycle period.\color{black} Indeed, based on the 20 ml/min mass flow rate used in the experiments of Ref.~\cite{zhu2022co2}, and the cross-sectional area of the unpacked DBD (circular ring), we can calculate a gas velocity of 0.1248 m/s and an advection time scale based on the reactor's length (35 mm) of 0.28 s, which is significantly larger than the AC period (0.5 ms).\color{black}  We note though the presence of the rods, might give rise to unsteady fluid dynamics phenomena (e.g. vortex shedding) that might influence the overall performance of the reactor and accumulation/convection of (mainly) neutral species in quasi-steady state. The gas temperature ($T_{g}$) is assumed constant at 300 K. This is a good approximation for the first AC cycles as energy transfer to the neutral species happends in longer time-scales (V-T, T-T relaxation). We note that in Ref.~\cite{ozkan2016influence}, $T_{g}$ was measured to around 500 K inside an empty reactor with a decreasing trend with decreasing AC frequency. As the main goal of this work is to investigate the plasma dynamics inside an AC cycle, this quite simplified chemistry set allows for an adequate description of the main charged species production, discharge evolution and spatiotemporal dynamics in the timescales studied here, while largely reducing the CPU burden of the simulation. 
Electrons and positive CO2 ions are initialized with a number density of $10^{9}$ $\mathrm{m^{-3}}$ while rest of ion and neutral species with a density of $10^{2}$ $\mathrm{m^{-3}}$. These values alsto correspond to the floor density for each species respectively. The initial/floor density of electron-$\mathrm{CO_{2}^{+}}$ pair are used to account for background ionization (e.g. from cosmic rays etc). The secondary electron emission (SEE) coefficient for the electrode boundary is set to $10^{-3}$ while for the dielectric rods and barrier to 1$\times$$10^{-2}$. The influence of these parameters on the discharge evolution falls outside the scope of this work. A further discussion on the choice of chemistry set and other parameters is given in Sec.~\ref{sec:discussion}.
%\begin{itemize}
%\item	4 species : electrons, negative ions, positive ions, neutral
%\item	5 reactions : ionization, two body and three body attachment, electron-ion and ion-ion recombination
%\item	Floor density for electrons and negative ions: $5\times10^{11}$ $m^{-3}$ 
%\item	Floor density for positive ions: $10^{12}$ $m^{-3}$ 
%\item	Secondary electron emission coefficient for electrode : 0.05
%\item	Secondary electron emission coefficient for dielectric : 0.01
%\item	Electric circuit resistance : 100 $\Omega$
%\end{itemize}
 
 \begin{table*}
\caption{\label{tabfour} Chemical reaction model used in the simulations. Rate units for two-body reactions in $\mathrm{m^{3}/s}$ and for three-body reactions in $\mathrm{m^{6}/s}$.} 
%\begin{indented}
\lineup
\begin{center}
\item[]\begin{tabular}{@{}*{3}{l}}
\br      
$\0\0\textbf{Species}$&\shortstack{electrons (e), $\mathrm{CO_{2}}$,  $\mathrm{CO_{2}^{+}}$, $\mathrm{O^{-}}$}\cr
$\0\0\textbf{}$&\shortstack{ $\mathrm{CO_{3}^{-}}$, $\mathrm{O_{2}^{-}}$,CO, O,  $\mathrm{O_{2}}$}\cr
\br                        
$\0\0\textbf{Reaction}$&\textbf{Rate constant}&\textbf{Reference}\cr
%$\textrm{Mesh support}$&$\textrm{Accuracy}$&\m$\textrm{LFA/LEA}$&\m$\textrm{Photoionization}$&$\textrm{\# of species and reactions}$&$\0\textrm{Parallelization}$\cr 
\mr
%$\0\0\textbf{Reactions}$&\shortstack{direct ionization, 2-body attachment,\\ 3-body attachment, \\ e-ion recombination, \\ion-ion recombination}\cr
%\mr
\0\0R1: e +  $\mathrm{CO_{2}}$ $\rightarrow$ $\mathrm{CO_{2}^{+}}$ + 2e &f(E) & Ref.~\cite{lisbon}$ ^{a}$\cr
\mr
\0\0R2: e +  $\mathrm{CO_{2}}$ $\rightarrow$ $\mathrm{CO}$ + $\mathrm{O}$ + e&f(E)&Ref.~\cite{lisbon}$ ^{a}$\cr
\mr
\0\0R3: e +  $\mathrm{CO_{2}}$ $\rightarrow$ $\mathrm{CO}$ + $\mathrm{O^{-}}$ &f(E)&Ref.~\cite{lisbon}$ ^{a}$\cr
\mr
\0\0R4: e +  $\mathrm{CO_{2}^{+}}$ $\rightarrow$ $\mathrm{CO}$ + $\mathrm{O}$ & $ 1.07\times 10^{-3}T_{e}^{-0.5}/T_{g} $&Ref.~\cite{wang2018modelling}$ ^{b}$\cr
\mr
\0\0R5: e +  $\mathrm{O_{2}}$ $\rightarrow$ $\mathrm{O}$ + $\mathrm{O}$ + e & $6.86\times 10^{-9}\mathrm{exp}(-6.29/T_{e})$ &Ref.~\cite{lieberman2005principles}$ ^{c}$\cr
\mr
\0\0R6: e +  $\mathrm{O_{2}}$ $\rightarrow$ $\mathrm{O}$ + $\mathrm{O^{-}}$&f(E)&Ref.~\cite{phelps}$ ^{a}$\cr
\mr
\0\0R7: e +  $\mathrm{O_{2}}$ + M $\rightarrow$ $\mathrm{O_{2}^{-}}$ + $\mathrm{M}$&f(E)&Ref.~\cite{phelps}$ ^{a}$\cr
\mr
\0\0R8: e + $\mathrm{O}$ + $\mathrm{M}$ $\rightarrow$ $\mathrm{O^{-}}$ + $\mathrm{M}$ &$1.0\times 10^{-31}$&Ref.~\cite{wang2018modelling}$ ^{d}$\cr
\mr
\0\0R9: $\mathrm{O^{-}}$ +  $\mathrm{CO_{2}}$ + $\mathrm{M}$$\rightarrow$ $\mathrm{M}$ + $\mathrm{CO_{3}^{-}}$ &$9\times10^{-29}$&Ref.~\cite{wang2018modelling}$ ^{d}$ \cr
\mr
\0\0R10: $\mathrm{CO_{3}^{-}}$ +  $\mathrm{CO_{2}^{+}}$ $\rightarrow$ 2$\mathrm{CO_{2}}$ + $\mathrm{O}$&$5\times10^{-7}$&Ref.~\cite{wang2018modelling}$ ^{b}$ \cr
\mr
\0\0R11: $\mathrm{O_{2}^{-}}$ +  $\mathrm{CO_{2}^{+}}$ $\rightarrow$  $\mathrm{CO}$  + $\mathrm{O_{2}}$ + $\mathrm{O}$&$6\times10^{-7}$&Ref.~\cite{wang2018modelling}$ ^{b}$ \cr
\mr
\0\0R12: $\mathrm{O}$  +$\mathrm{CO}$ +  $\mathrm{M}$$\rightarrow$ $\mathrm{CO_{2}}$ + $\mathrm{M}$ & $1.7\times 10^{-33}\mathrm{exp}(-1510/T_{g})$ &Ref.~\cite{aerts2012influence}$ ^{d}$\cr
\mr
\0\0R13: $\mathrm{O}$ +$\mathrm{O}$ +  $\mathrm{M}$ $\rightarrow$ $\mathrm{O_{2}}$ + $\mathrm{M}$& $1.27\times 10^{-32}(298/T_{g})\mathrm{exp}(-170/T_{g})$ &Ref.~\cite{wang2018modelling}$ ^{d}$\cr
\mr
\0\0R14: $\mathrm{CO}$ +$\mathrm{CO_{3}^{-}}$ $\rightarrow$ 2$\mathrm{CO_{2}}$ + e  & $5\times 10^{13}$ &Ref.~\cite{ono1984negative}$ ^{b}$\cr
%\0\02D/2.5D&Structured/Unstructured/Hybrid  &2nd order (space and time)&Both&Yes &\0Arbitrary (tabulated)&\0Yes (MPI)\cr
\br
\0\0$ ^{a}$  Bolsig+.& \cr
\0\0$ ^{b}$  Units of $\mathrm{cm^{3}/s}$. $\mathrm{T_{e}}$ and $\mathrm{T_{g}}$ in K. \cr
\0\0$ ^{c}$  Units of $\mathrm{cm^{3}/s}$. $\mathrm{T_{e}}$ in eV. \cr
\0\0$ ^{d}$  Units of $\mathrm{cm^{6}/s}$. M=$\mathrm{CO_{2}}$. $\mathrm{T_{g}}$ in K.  \cr
%\0\0$ ^{3}$ XXX& \cr
\end{tabular}
\end{center}
%\end{indented}
\end{table*}

%\begin{table}
%\caption{\label{tabthree} Chemistry and plasma related parametres used in the simulations.} 
%\begin{indented}
%\lineup
%\item[]\begin{tabular}{@{}*{2}{l}}
%\br                              
%$\0\0\textbf{Species}$&\shortstack{electrons (e), $CO_{2}$, \\ $CO_{2}^{+}$, CO, $O^{-}$}\cr
%%$\textrm{Mesh support}$&$\textrm{Accuracy}$&\m$\textrm{LFA/LEA}$&\m$\textrm{Photoionization}$&$\textrm{\# of species and reactions}$&$\0\textrm{Parallelization}$\cr 
%\mr
%%$\0\0\textbf{Reactions}$&\shortstack{direct ionization, 2-body attachment,\\ 3-body attachment, \\ e-ion recombination, \\ion-ion recombination}\cr
%%\mr
%$\0\0\textbf{Floor density for e and  $CO_{2}^{+}$}$&$1\times10^{9}$ $m^{-3}$\cr
%\mr
%$\0\0\textbf{Floor density for rest of species}$&$10^{2}$ $m^{-3}$\cr
%\mr
%$\0\0\textbf{SEE coef. for electrode}$&0.0001\cr
%\mr
%$\0\0\textbf{SEE coef. for dielectric layer and rods}$&0.05\cr
%%\0\02D/2.5D&Structured/Unstructured/Hybrid  &2nd order (space and time)&Both&Yes &\0Arbitrary (tabulated)&\0Yes (MPI)\cr
%\br
%\end{tabular}
%\end{indented}
%\end{table}

\section{Numerical results}\label{sec:results}

\subsection{Electrical characteristics}\label{subsec:electric}
\begin{figure}[h]
\centering
\includegraphics[width=1\linewidth]{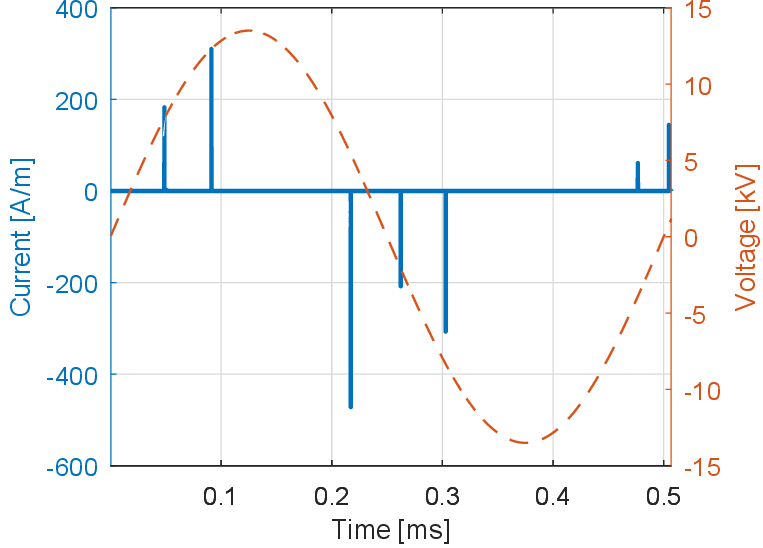} 
\caption{\label{fig:IV_6rods} \color{black} Discharge current (I) - voltage (I-V) characteristics in the first AC cycle. Current values are extrapolated to the total 2D reactor volume. \color{black}
}
\end{figure}

%\color{red}HERE MAKE SURE THAT ALL NUMBERS ETC CORRESPOND TO THE NEW I-V GRAPHS FROM THE 12 RODS CASE!!!\color{black}
The discharge (conduction) current - applied voltage curve of the discharge and during the slightly extended first AC period, is shown in Fig.~\ref{fig:IV_6rods}. The discharge current corresponds to the full reactor geometry (24 times the discharge current obtained in the simulation domain) and as such we assume that identical (both in time occurence and intensity) MDs occur inside each rod region. In total, six (6) MDs ignite inside the first full AC cycle, three of those carrying positive current and two carrying negative current. We note here that we follow the convention of Ref.~\cite{kogelschatz2003dielectric}, where a streamer (cathode or anode directed) discharge is denoted as a propagating ionization front which transits to a MD upon impact to the dielectric (or electrode) surface (other authors use the term filamentary discharge). The last current peak visible, occurs at the very beginning of the second AC cycle. Peak MD currents range from approximately 70 to 480 A/m. In Ref.~\cite{ozkan2016influence}, the authors measured peak MD currents in the order of 10 - 200 mA for an empty DBD reactor operating in $\mathrm{CO_{2}}$ with a discharge length of 100 mm. Assuming a uniform distribution in the z-direction (a rough assumption), the experimentally measured MD currents can be normalized to 0.1 - 2 A/m.  Taking into account the fact that each MD in our model is a superposition of several MDs occuring in the azimuthal direction (as we assume symmetric/periodic boundary conditions), we estimate a single MD peak current calculated from our model to be in the 3 - 20 A/m range (dividing the total current by the 24 factor). Considering the differences in applied voltage, frequency, discharge gap, dielectric materials used, empty reactor setup of the aforementioned reference and the assumption of discharge uniformity in the z-direction, the calculated MD current values of this work, are close to these experimental findings.

Looking in more detail on  each MD phase, we note that the first MD occurs during the positive half cycle at approximately t$=$48.7 $\mathrm{\mu}$s when the applied voltage is around 7.75 kV.  A second MD appears at approx. t$=$91.21 $\mathrm{\mu}$s when the applied voltage is 12.3 kV and still at the rising phase of the positive half-cycle. The time interval between these two positive current-carrying MDs is approx. 42.5 $\mathrm{\mu}$s. As we will see in the detailed analysis of each MD (see below), this time interval is not linked to the relaxation and decay of a single MD but mainly to the redistribution of the electric field due to space/surface charge and consequent breakdown at different regions of the reactor. During the remaining of the positive half-cycle rising phase, no MD is present. During the falling phase of the positive half-cycle, and when the applied voltage is around 5.4 kV, a third MD ignites at t$=$217.2 $\mathrm{\mu}$s carrying negative current. Approximately 45 $\mathrm{\mu}$s later and while the applied voltage is now at the rising phase of the negative half-cycle, a fourth MD occurs at approx. t$=$262.3 $\mathrm{\mu}$s (applied voltage of -2.1 kV). The time interval between these negative current-carrying MDs is similar to the first set of positive MDs. The fifth MD which ignites at approx. t$=$303 $\mathrm{\mu}$s (applied voltage of -8.34 kV), is delayed compared to the time-interval of the 1st set of negative (N) MDs as it occurs 86 $\mathrm{\mu}$s later than the 2nd N-MD. Similarly to the positive half-cycle, no N-MDs are present at the remaining of the negative half-cycle. Nevertheless, memory effects lead to to the appearance of a P-MD at approx. t$=$477 $\mathrm{\mu}$s (applied voltage of -3.95 kV), which marks the initiation of the second positive discharge phase. We finally note the additional P-MD igniting at the very beginning of the second AC cycle, whose dynamics fall outside the scope of the current work.

%\color{red}HERE MAKE SURE THAT ALL NUMBERS ETC FROM THE EXPERIM PAPER CORRESPOND TO THE 12 RODS CASE!!!\color{black}

\begin{figure}[h]
\centering
\includegraphics[width=1\linewidth]{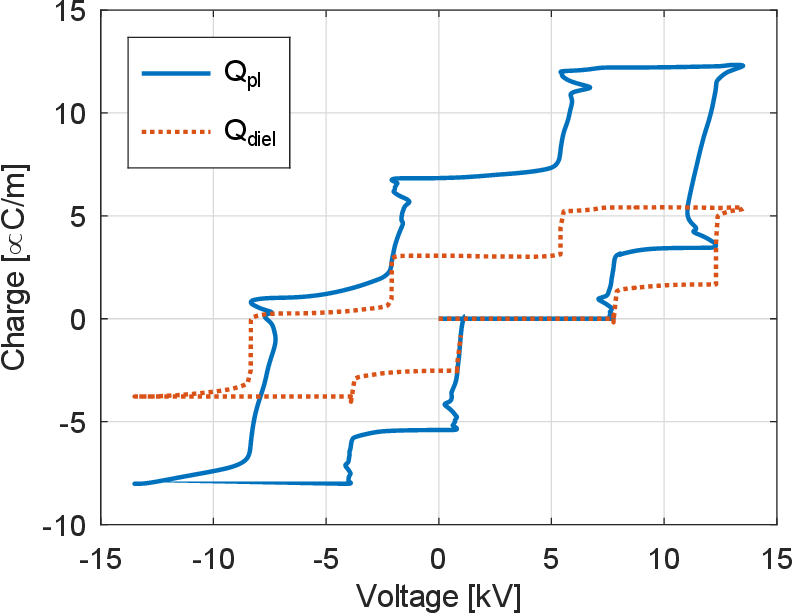} 
\caption{\label{fig:lissajous} \color{black} Lissajous (Q-V) curve in the first AC cycle. Charge values are extrapolated to the total 2D reactor volume and correspond to the plasma charge ($Q_{pl}$) and the total deposited charge in all dielectric surfaces (layer and rods, ($Q_{diel}$)).\color{black}
}
\end{figure}

In Fig.~\ref{fig:lissajous}, the calculated Lissajous plot (Q-V diagram, where $Q=Q_{pl}$, is the total plasma charge and  $Q=Q_{diel}$, is the total charge deposited in the dielectric layer and rods in the full 2D reactor volume) is shown. $Q_{pl}$, is calculated by integrating the (full reactor) discharge current over time. $Q_{diel}$, is calculated by integrating the (full reactor) surface charge density($\mathrm{C/m^{2}}$) over all dielectric boundary lines. Note that the time integrals correspond to the slightly extended first AC cycle which explains the overlap observed in the 0-1 kV rising voltage range. In typical Q-V diagrams related to DBDs, the diagram has a parallelogram shape where the on and off (capacitive) phases of the plasma discharge are clearly visible~\cite{peeters2019electrical}. Due to the presence of MDs in both positive and negative AC half-cycles, the Q-V diagram presents a stepped character. The small oscillations of the $Q_{pl}$ curve at the end of each step are due to the presence of the ballast resistor whose influence kicks in when the discharge current is large. The positive and negative half-period total charge is 6.8 $\mathrm{\mu}$C/m and 6.06 $\mathrm{\mu}$C/m respectively, which scaled with the reactor/discharge length (35 mm) translates to 0.24 and 0.21 $\mathrm{\mu}$C respectively. The average discharge (absorbed) power in the first AC cycle can be easily calculated by the time-average integral of the voltage ($V$) - current ($I$) product inside the first AC cycle ($\mathrm{P_{avg}=f*\int_{0}^{T}{V\cdot I\:dt}}$, where $T$ is the AC cycle period), which gives us a value of $\mathrm{P_{avg}}$=353.42 W/m. The same value is obtained by integrating the Lissajous plot (time averaged integral of voltage over charge). Considering the discharge length (35 mm) and assuming discharge uniformity in the z-direction (out of plane), we estimate the average discharge power to be near 12 W under the studied conditions. In Ref.~\cite{zhu2022co2}, the discharge power for the same packing number and materials is close to 26 W, but the authors do not clarify if the applied voltage of 27 kV translates to peak-to-peak or peak voltage. In addition, a direct comparison is not adequate here as our simulations reports only the first AC-cycle time-averaged values.  Moreover in the same reference, the authors report the transferred discharge charge (Lissajous plots) which roughly lies in the -0.75  to 0.75  $\mathrm{\mu}$C range, for 24 (1 mm) $\mathrm{ZrO_{2}}$ rods (fitting curves not takinto into account the instantaneous steps due to each MD). Taking into account the experimental reactor (discharge) length of 35 mm, we can rescale plot~\ref{fig:lissajous} to obtain values which lie in the -0.3 to 0.5 $\mathrm{\mu}$C range, which is close to the experimental findings considering that we're focusing only on the first AC cycle here, we assume uniformity in the z-direction and the rod packing number of this study is 12 instead of 24. The charge deposited on the dielectric surfaces, $Q_{diel}$, (both dielectric layer and rods) is significant and compares with the plasma charge which indicates the important effects of surface charging on the operation of the PB-DBD. 
%Lastly, a few words on the number of MDs per half-period: In Ref.~\cite{zhu2022co2}, the authors state that the number of MDs per half period for the empty reactor is 27 while for the same packing material, density and operating conditions as in our simulations this number is doubled to 54. In the numerical results and due to the 2D approximation, it is impossible to reproduce, let alone compare, the number of MDs inside the half-cycle in the three-dimensional reactor. In the axial (z-) dimension, MDs should form rather chaotically in time and space due to non-uniform dielectric charging of the rods, electrostatic interference between neighboring MDs, gas flow effects, edge effects and other phenomena that can not be captured within our model. With a very simplistic extrapolation and assuming a) a MD depth in the z-direction of approximately 1 mm, b) a spatial interval of 1 mm between neighboring MDs in the z-direction and c) the reactor length of 35 mm, more than 50 MDs could appear in the half-cycle - an estimate close to the experimental observations but once again quite crude. Note than an increase in discharge power, results to an increase in the number (and intensity) of MDs inside a half-period~\cite{ozkan2016influence}.  %A better comparison to experimental findings can be made with the Lissajous curve as plotted in Fig~\ref{lissajous_6rods}. 

In the next section, we focus on each MD pulse in order to understand the individual discharge ignition, development and decay/afterglow phases.

\subsection{Discharge development during each current pulse}\label{subsec:pulses}
\subsubsection{First current pulse and its afterglow}\label{subsubsec:1stpulse}
\begin{figure}[h]
\centering
\includegraphics[width=1\linewidth]{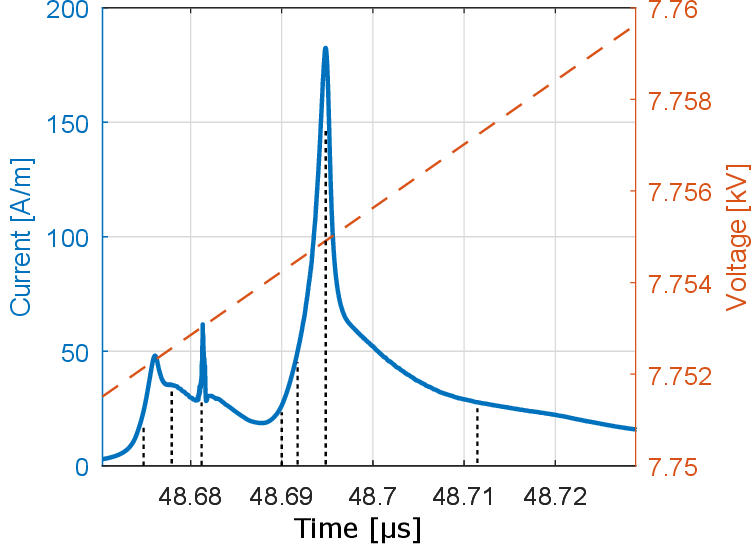} 
\caption{\label{fig:IV_1stPulse} \color{black} Current [A/m] and applied voltage [V] during the 1st MD. The vertical dashed lines indicate the time instants used for subsequent plots on discharge characteristics. \color{black}
}
\end{figure}

\begin{figure}[h]
\centering
\includegraphics[width=1\linewidth]{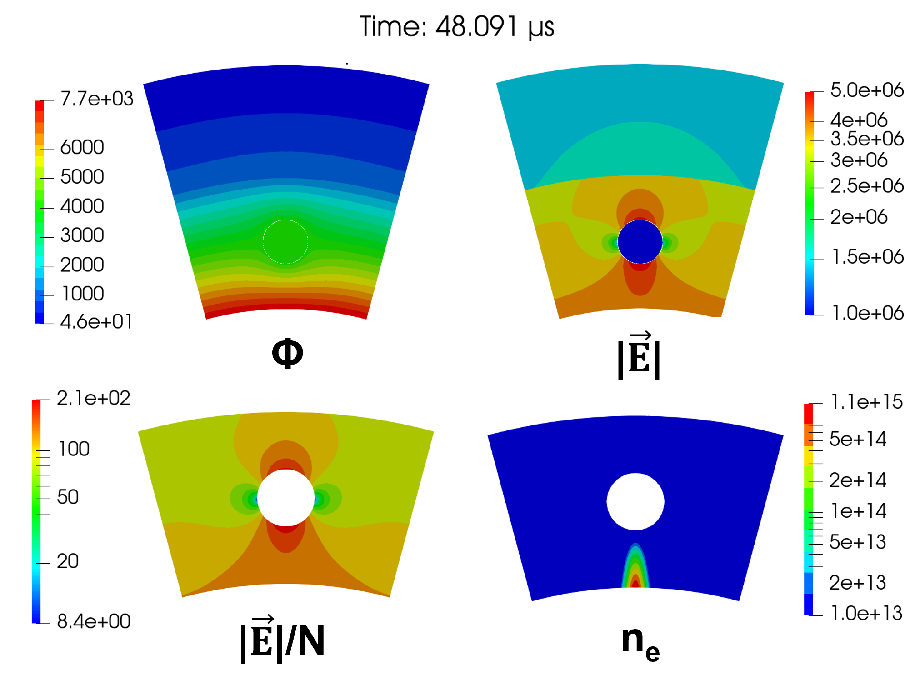} 
\caption{\label{fig:E_1stPulse_0} \color{black} Potential $\Phi$ [V], electric field magnitude $|\vec{E}|$ [V/m] (log-scale), reduced electric field $|\vec{E}/N|$ [Td] (log-scale) and electron density, $n_{e}$, [$\mathrm{m^{-3}}$] (log-scale, min values scaled to $\mathrm{10^{13}}$ $\mathrm{m^{-3}}$ ) contours, just before the 1st MD. The position of the dielectric rod is indicated with the white dashed circle.\color{black}
}
\end{figure}

Fig.~\ref{fig:IV_1stPulse} depicts the I-V curve during the 1st MD, which consists of three distinct current peaks. Initially and before breakdown, the geometry of the coaxial PB-DBD enhances the electric field in the region between the inner electrode, the rods and the dielectric surface, where initial ionization takes place as depicted in Fig.~\ref{fig:E_1stPulse_0} at t$=$48.091 $\mathrm{\mu}$s. Therein, and especially near the poles of the dielectric rod (due to polarization effects) the reduced electric field ($E/N$, where $N$ is the neutral gas density) reaches values of approx. 200 Td, obtaining larger vales at the inner pole region (inner here denotes the region between the innner electrode and the rods). Due to the large drift velocity of electrons, the initial electron avalanche moves towards and accumulate at the inner electrode (anode) leaving behind a positive ion cloud (density of $\mathrm{10^{17}}$ $\mathrm{m^{-3}}$, not shown here) which fills in the inner electrode-rod-dielectric region. The first and second current peaks correspond to a glow-to-streamer-to-MD discharge transition between the inner electrode and the dielectric rods.  A quasi-neutral glow discharge is initially formed near the inner electrode region (first current peak). In subsequent times, the positive space charge and volumetric charge separation locally distorts the potential distribution, enhancing the electric field above the inner electrode. The space charge induced electric field enhancement does not occur close to the inner electrode (as typically found in point-to-plane configurations with pulsed excitation) but in the volumetric region closer to the dielectric rod (around 2/3 of the electrode-rod gap distance). This is a particular characteristic of streamer initiation in AC-DBDs compared to pulsed DBDs (or even needle-plane configurations), and agrees well with positive streamer initiation in AC surface DBDs where the streamer initiates at a distance from the exposed electrode due to the long AC time-scales and the development of a charge separation region~\cite{kourtzanidis2020self}. The initial glow-like discharge quickly transits to a rather thin (thickness of approx. 0.1 mm) self-propagating (positive/cathode-directed) streamer in the form of an ionization wave which travels towards the dielectric rod and bringes the gap in less than 5 ns (second current peak). 

\begin{figure}[h]
\centering
\includegraphics[width=1\linewidth]{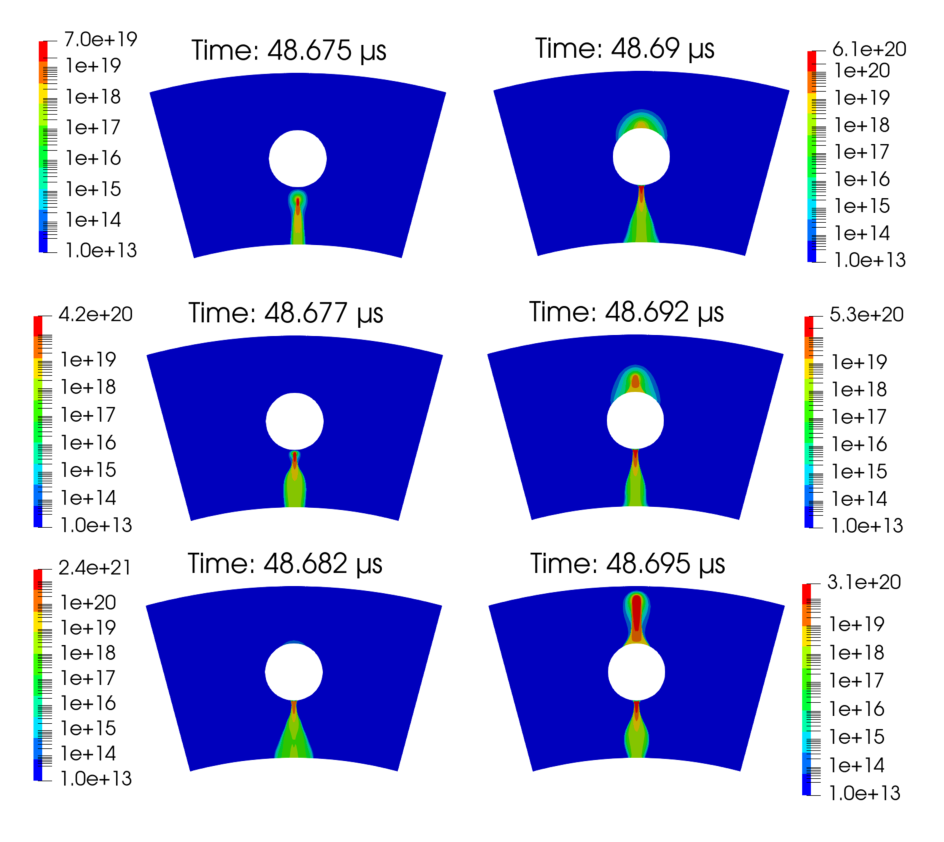} 
\caption{\label{fig:ne_1stPulse} \color{black} Electron density contours (units of $\mathrm{m^{-3}}$, log-scale, min values scaled to $\mathrm{10^{13}}$ $\mathrm{m^{-3}}$) at different time-instants (see vertical dashed lines at Fig.~\ref{fig:IV_1stPulse}), during the 1st MD.  \color{black}
}
\end{figure}

 Fig.~\ref{fig:ne_1stPulse} shows the electron density contours at different time-instants, during the MD development. The quasineutral MD sustains electron densities in the order of $\mathrm{10^{20}-10^{21}}$ $\mathrm{m^{-3}}$ inside the streamer body and $\mathrm{10^{17}-10^{19}}$ $\mathrm{m^{-3}}$ inside the glow region. A very thin (thickness of 10-20 $\mathrm{\mu}$m in agreement with previous studies~\cite{steinle1999two}) cathode-like layer is formed at the dielectric rod which sustains a positive ion charge and very high electric fields, allowing for relatively efficient positive charging of the dielectric rod surface. The positive dielectric rod charging eventually chokes the MD due to the redistribution (reduction) of the electric field in this region which starts decaying (reducing the current flow). At this point in time, the potential inside the whole PB-DBD reactor has been heavily disturbed: the initial streamer/MD itself holds a positive potential (due to the high positive space charge around its head) and even though the electric field inside its quasi-neutral body is greatly reduced, it effectively plays a role of virtual anode which leads to an effective reduction of the reactor's gap at the rod positions. This effect which adds up to the positive surface charging of the dielectric rod inner side, greatly enhances the electric field in the outer electrode-rod-dielectric region (outer here refers to the region between the rod and the dielectric layer), especially near the rod's outer pole. A second streamer discharge is initiated at this region, very close to the dielectric rod outer pole (due to the quite high reduced electric fields at this region), propagating towards the dielectric layer and bridging the gap in less than 15 ns (third current peak). Electron densities are of the same order of magnitude as inside the 1st streamer body but the streamer thickness is almost three times larger (approx. 0.3 mm). Upon impingement on to the dielectric layer and similar to the 1st streamer, a cathode layer forms but due to the lower permittivity of the dielectric layer (compared to the dielectric rods), its thickness is significant larger (approx. 50 $\mathrm{\mu}$m). Therein the electric field is sustained to values higher than 20 MV/m, while a positive space charge density of several (1-9) $\mathrm{C/m^{-3}}$ is present (electrons are quickly lost). The MD starts to transit to a Surface Ionization Wave (SIW or surface streamer) as evidenced by its slight expansion on the dielectric surface on both azimuthal directions, but the SIW propagation is very limited due to the relative low E/N values (around 500 Td) at the SIW heads. 

\begin{figure}[h]
\centering
\includegraphics[width=1\linewidth]{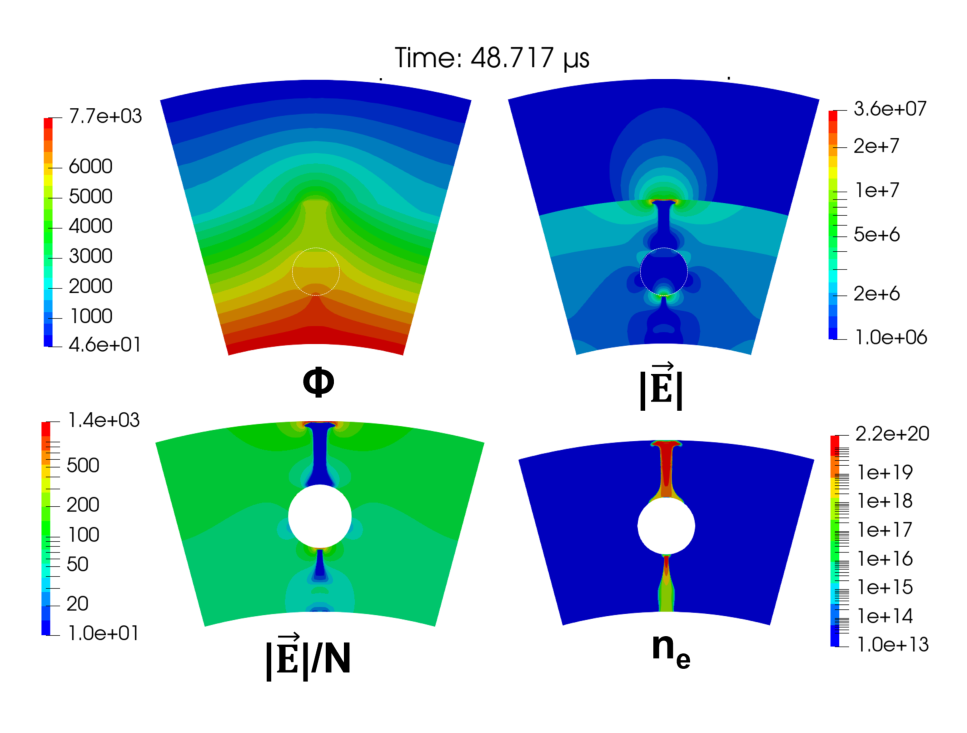} 
\caption{\label{fig:E_1stPulse} \color{black} Potential $\Phi$ [V], electric field magnitude $|\vec{E}|$ [V/m] (log-scale), reduced electric field $|\vec{E}/N|$ [Td] (log-scale,  min value scaled to10 Td for visualization purposes) and electron density, $n_{e}$, [$\mathrm{m^{-3}}$] (log-scale, min values scaled to $\mathrm{10^{13}}$ $\mathrm{m^{-3}}$ ) contours, at the decaying phase of the 1st  MD. The position of the dielectric rod is indicated with the white dashed circle.\color{black}
}
\end{figure}

\begin{figure}[h]
\centering
\includegraphics[width=1\linewidth]{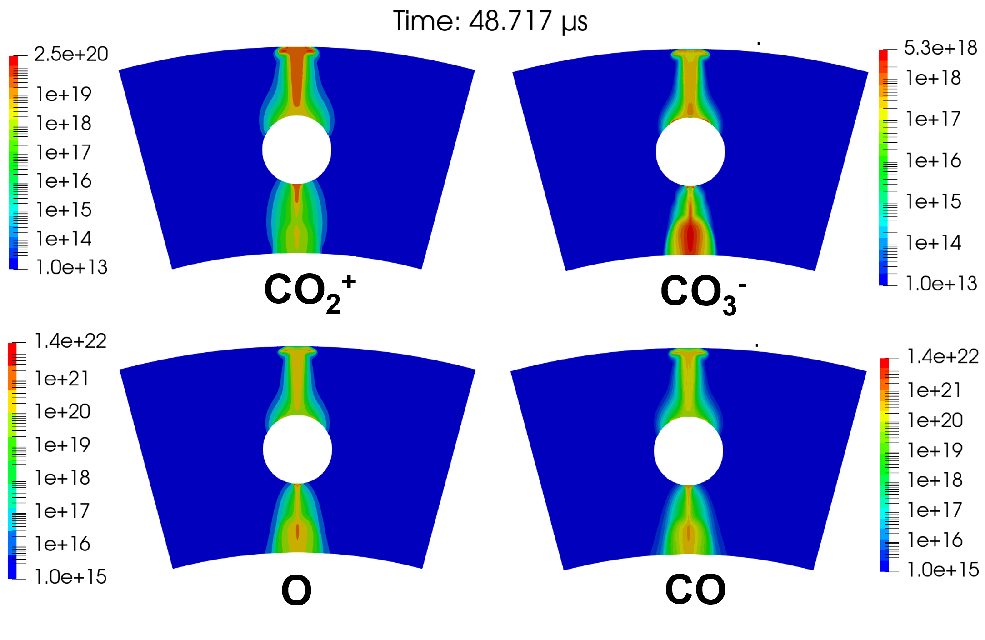} 
\caption{\label{fig:species_1stPulse} \color{black} Major species density contours  (units of $\mathrm{m^{-3}}$, log-scale, min values scaled to $\mathrm{10^{13}}$ $\mathrm{m^{-3}}$ for charged species and $\mathrm{10^{15}}$ $\mathrm{m^{-3}}$ for neutral species)  at the decaying phase of the 1st MD.\color{black}
}
\end{figure}

During the decaying phase of the current pulse at t$=$48.717 $\mathrm{\mu}$s, the reduced electric field inside the MD zone has collapsed (non-uniformly) sustaining a positive potential as shown in Fig.~\ref{fig:E_1stPulse}. The potential redistribution due to the presence of the MD leads to zones of enhanced reduced fields (several 100 Td) surrounding both inner and outer streamer discharges, while the reduced field at the MD/dielectric layer interface reaches values of 1400 Td. At this time instant, the charged and neutral species distribution follow the MD development with particular attributes per species, as shown in Fig.~\ref{fig:species_1stPulse}. $\mathrm{CO_{2}^{+}}$ ions are preferentially formed inside the streamers' bodies with densities in the order of $\mathrm{10^{19}-10^{20}}$ $\mathrm{m^{-3}}$. The dominant ion species calculated by our model are found to be $\mathrm{CO_{3}^{-}}$ ions (due to the rapid three-body transformation of $\mathrm{O{-}}$ ions - R9), whose density is higher inside the glow region of the inner discharge (order of  $\mathrm{10^{18}}$ $\mathrm{m^{-3}}$) while their density inside the streamers bodies are one order of magnitude lower ($\mathrm{10^{17}}$ $\mathrm{m^{-3}}$). Compared to the electron distribution, we note that the low ion drift velocity results to a more diffusive distribution (electrons are quickly accelerated and lost to the surfaces when diffused away from the quasi-neutral MD regions). The main neutral species produced by the MD are O atoms and CO molecules, both having average densities in the order of $\mathrm{10^{20}}$ $\mathrm{m^{-3}}$. Interestingly,  $\mathrm{O_{2}^{-}}$ ion densities are negligible (lower than $\mathrm{10^{11}-10^{12}}$ $\mathrm{m^{-3}}$).

\begin{figure}[h]
\centering
\includegraphics[width=1\linewidth]{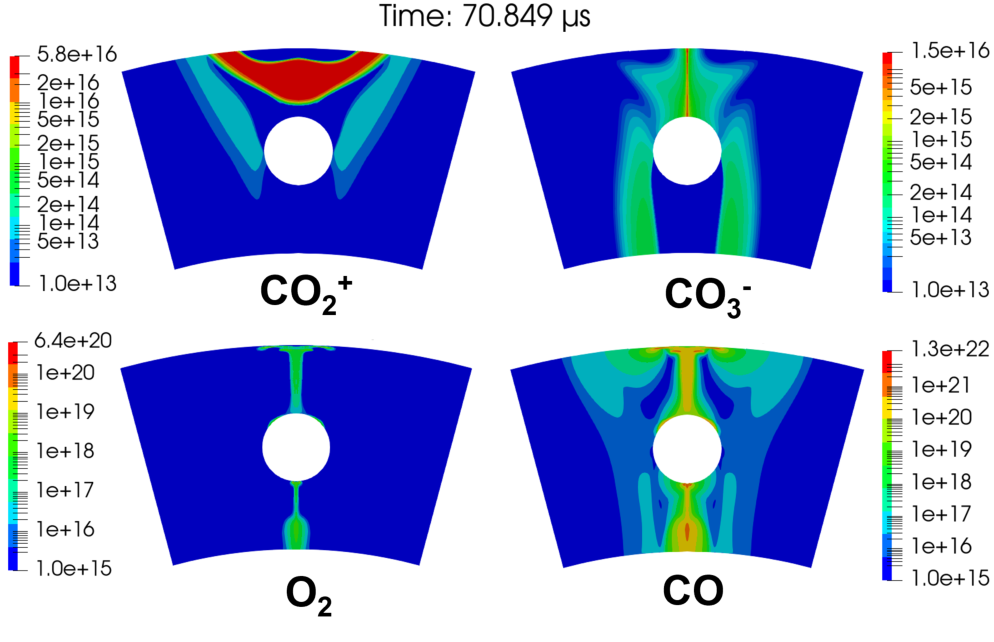} 
\caption{\label{fig:species_1stPulse_long} \color{black} Major species density contours  (units of $\mathrm{m^{-3}}$, log-scale, min values scaled to $\mathrm{10^{13}}$ $\mathrm{m^{-3}}$ for charged species and $\mathrm{10^{15}}$ $\mathrm{m^{-3}}$ for neutral species)  at the long-term afterglow phase of the 1st  MD. O atoms (not shown here) present very similar profiles to the CO molecules. \color{black}
}
\end{figure}

\begin{figure}[h]
\centering
\includegraphics[width=1\linewidth]{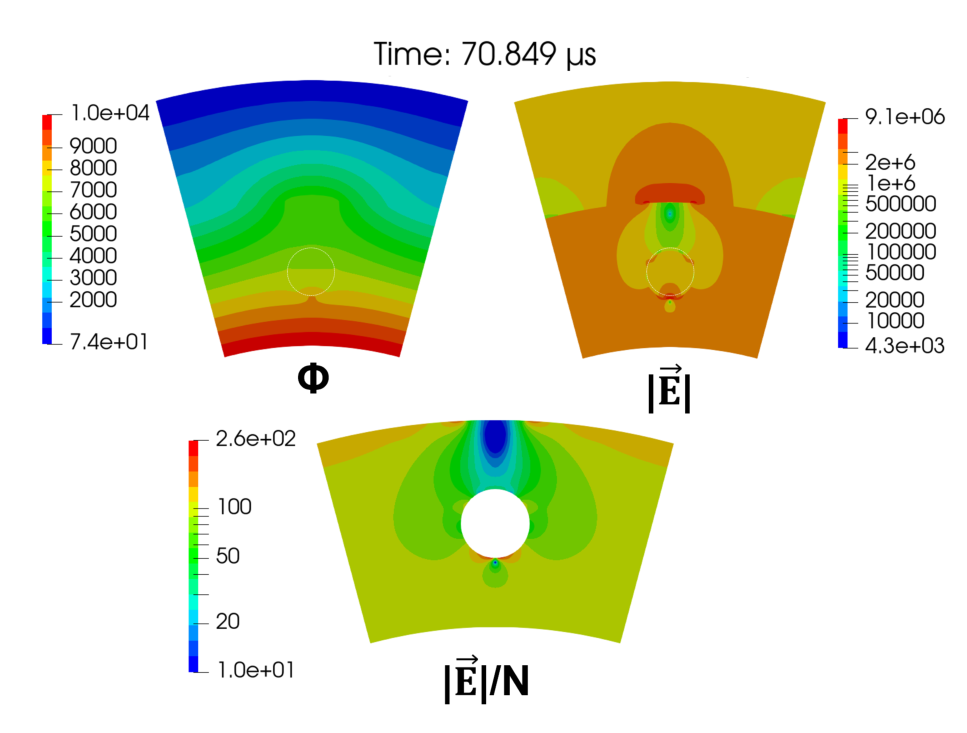} 
\caption{\label{fig:E_1stPulse_long} \color{black} Potential $\Phi$ [V], electric field magnitude $|\vec{E}|$ [V/m] (log-scale) and reduced electric field $|\vec{E}/N|$ [Td] (log-scale,  min values scaled to 10 Td for visualization purposes) contours, at the long-term afterglow phase of the 1st MD. The position of the dielectric rod is indicated with the white dashed circle.\color{black}
}
\end{figure}

The time-period until the fomation of the next current pulse, are here defined as the short-term afterglow and long-term afterglow phases. During the short-term afterglow (lasting approximately 1 $\mathrm{\mu}$s), electron diffusion and recombination losses dominate and the quasi-neutral plasma density starts decaying. This phase is extremely important for the operation of the DBD discharge (and applicable to any type of DBD reactor or surface DBD actuators) as it dictates the dielectric charging intensity and spatial distribution: as the plasma density (and its conductivity) decays, the electric field inside the plasma body grows allowing the (high density) charged species to escape the plasma body. These charged species will drift due to the electric field, with high velocities near the high-electric field regions surrounding the plasma and especially near the plasma-dielectric regions. Positive ions will then start charging the dielectric rod and dielectric layer, adjacent to the MD. When then induced potential due to the surface charges become sufficiently positive, these ion drift (and consequent surface charging) will jump to a downstream location on the dielectric, extending the surface charged region way beyond the MD thickness. 
 In later time-instants and despite the continously rising, positive, applied voltage, electron diffusion and recombination losses continue to dominate, so that the electron density is gradually diminished on a $\mathrm{\mu}$s time-scale.  We define here, the long-term afterglow time-instant, as the instant after a current pulse when the electron density has dropped below $\mathrm{10^{12}}$ $\mathrm{m^{-3}}$ so that electrons are effectively evacuated from the reactor. As so, the long-term afterglow of major ion and neutral species at  t$=$70.849 $\mathrm{\mu}$s (approx. 22 $\mathrm{\mu}$s after the current pulse) is plotted in Fig.~\ref{fig:species_1stPulse_long}. At the same time-instant, in Fig.~\ref{fig:E_1stPulse_long}, we plot the potential distribution, electric field magnitude and reduced electric field. We first note that in Fig.~\ref{fig:E_1stPulse_long},  the dielectric charging memory effects are clearly visible especially on the dielectric layer zone. As noted above, the charged portion of the dielectric is significantly larger than the MD length, and it approaches the rod diameter (approx. 1 mm). The redistributed potential results to reduced electric fields lower than 20 Td in the outer electrode-rod-dielectric region, and around 100 Td in the rest of the reactor. Both dominant charged species have heavily decayed at this point (note again the ion distribution which follows the uncharged portion of the dielectric surface). The dominant neutral species remain O atoms (not shown as their spatial distribution and min-max densities are very similar to CO) and CO molecules, whose maximum densities remain relatively unchanged but their population is now expanded in space due to both electron impact and secondary chemical reactions following the zones of ion/electron drift. We note here that the neutral species diffusion length scale inside a full AC period (0.5 ms) is approximately 0.1 - 0.15 mm and thus diffusion plays a very limited role in the time-scales considered in this work. The population of $\mathrm{O_{2}}$ molecules has increased significantly during the afterglow phase and inside the discharge body, reaching values higher than $\mathrm{10^{17}}$ $\mathrm{m^{-3}}$ due to R13.

 In the next subsection, we focus on the second current pulse and the discharge/species spatiotemporal development.

 %%%%%%%%%%%%%% 2nd current pulse
 \subsubsection{Second current pulse and its afterglow}\label{subsubsec:2ndpulse}
\begin{figure}[h]
\centering
\includegraphics[width=1\linewidth]{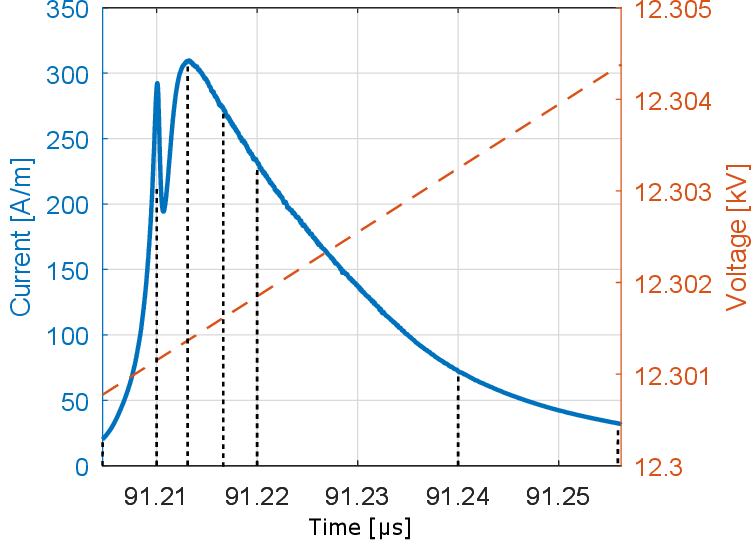} 
\caption{\label{fig:IV_2ndPulse} \color{black} Current [A/m] and applied voltage [V] during the 2nd MD. \color{black}
}
\end{figure}

 Fig.~\ref{fig:IV_2ndPulse} depicts the I-V curve during the second MD, which consists of two distinct current peaks in the 300 A/m range. 
 
\begin{figure}[h]
\centering
\includegraphics[width=1\linewidth]{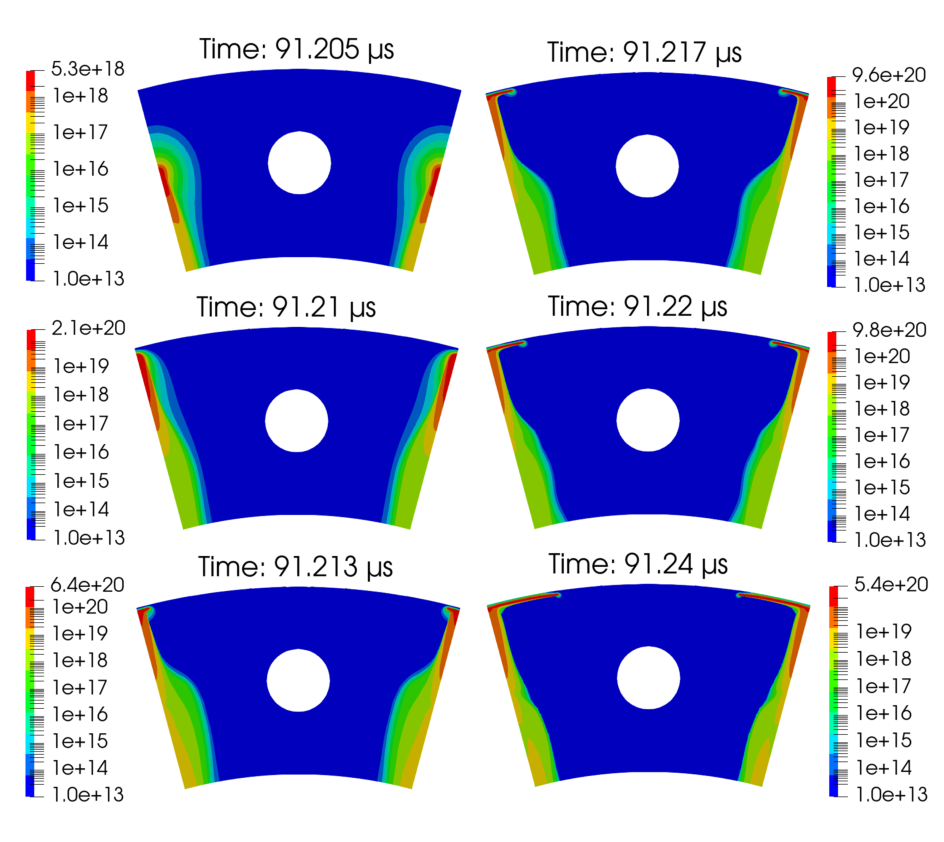} 
\caption{\label{fig:ne_2ndPulse} \color{black} Electron density contours (units of $\mathrm{m^{-3}}$, log-scale, min values scaled to $\mathrm{10^{13}}$ $\mathrm{m^{-3}}$) at different time-instants (see vertical dashed lines at Fig.~\ref{fig:IV_1stPulse}), during the 2nd MD.  \color{black}
}
\end{figure}

 Just before breakdown, (e.g. at t$=$90.55 $\mathrm{\mu}$s), remnant and persisting surface charges from the first positive pulse (as we've seen in Fig.~\ref{fig:E_1stPulse_long}), favor ionization in the azimuthal position between the dielectric rods, in contrast to the first current pulse development. As seen in Fig.~\ref{fig:ne_2ndPulse} (electron density contours at different time-instants, during the MD development), the initial glow discharge, transits to a streamer with a starting point near the centerline of the reactor. In less than 5 ns, the streamer reaches the dielectric layer and transforms to a surface ionization wave (SIW) which propagates along the dielectric surface in both positive and negative azimuthal directions. The SIW (or surface streamer) is detached from the dielectric forming a very thin sheath layer. We have already remarked this important feature of SIWs in surface DBDs operating in atmospheric air, which requires both a very fine mesh at such regions and accurate time-integration schemes which respect the dielectric relaxation time to be captured~\cite{kourtzanidis2020self, kourtzanidis2021electrohydrodynamic}. The surface streamer stops propagating at approx. 91.24 $\mathrm{\mu}$s, covering a total arc length, $s$, of approx. 2.6 mm. The arc length here is calculated by  $s=\pi r_{diel}\theta_{SIW}/180$, where $r_{diel}$=20 mm and $\theta_{SIW}$ is the central angle of the arc formed by the SIW (approx. $\mathrm{15^{\circ}}$). At this time-instant, the electric field magnitude at the SIW head is not sufficient to sustain its propagation. Electron densities inside the volume streamer and SIW body are in the order of  $\mathrm{10^{20}}$ $\mathrm{m^{-3}}$, and $\mathrm{10^{18}}$ $\mathrm{m^{-3}}$ in the glow region.

\begin{figure}[h]
\centering
\includegraphics[width=1\linewidth]{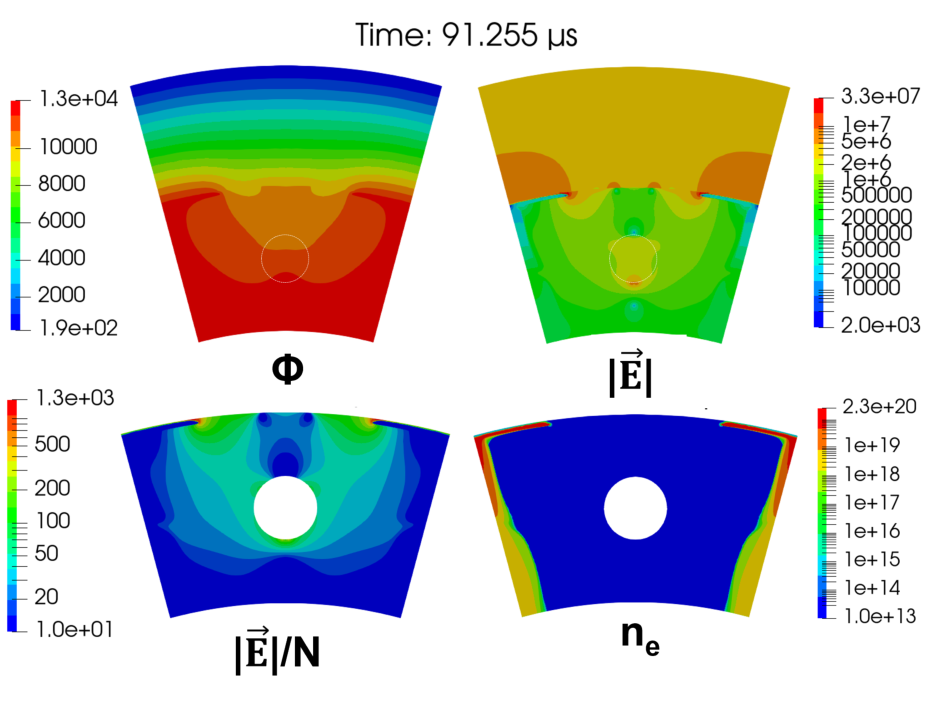} 
\caption{\label{fig:E_2ndPulse} \color{black} \color{black} Potential $\Phi$ [V], electric field magnitude $|\vec{E}|$ [V/m] (log-scale), reduced electric field $|\vec{E}/N|$ [Td] (log-scale,  min values of 10 Td for visualization purposes) and electron density, $n_{e}$, [$\mathrm{m^{-3}}$] (log-scale, min values scaled to $\mathrm{10^{13}}$ $\mathrm{m^{-3}}$ ) contours, at the decaying phase of the 2nd  MD. The position of the dielectric rod is indicated with the white dashed circle.\color{black}
}
\end{figure}

 During the decaying phase of the current pulse at t$=$91.255 $\mathrm{\mu}$s, the reduced electric field inside the MD zones which mainly correspond to the streamer and SIW, has collapsed  sustaining a positive potential as shown in Fig.~\ref{fig:E_2ndPulse} and similarly to the 1st pulse. Reduced fields at the SIW head reach values around 1300 Td while inside the plasma regions fall below 10 Td. The positive potential that is sustained by the SIW which acts as a virtual positively-biased electrode, leads to the zones of enhanced (more than 100 Td) $E/N$ shown in the same figure.     
 
\begin{figure}[h]
\centering
\includegraphics[width=1\linewidth]{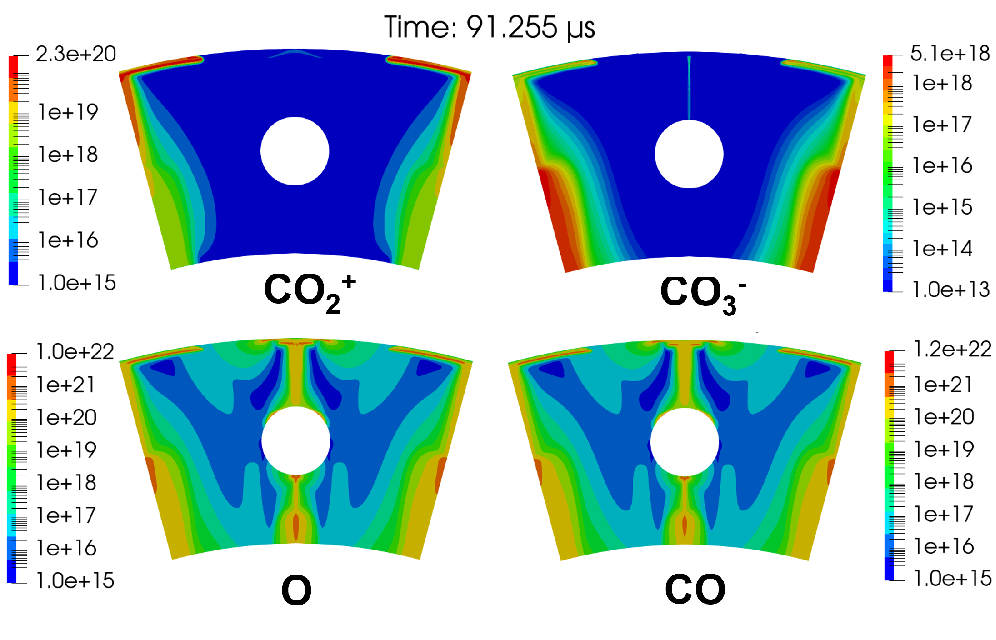} 
\caption{\label{fig:species_2ndPulse} \color{black} Major species density contours  (units of $\mathrm{m^{-3}}$, log-scale, min values scaled to $\mathrm{10^{13}}$ $\mathrm{m^{-3}}$ for charged species and $\mathrm{10^{15}}$ $\mathrm{m^{-3}}$ for neutral species)  at the decaying phase phase of the 2nd  MD.\color{black}
}
\end{figure}

\begin{figure}[h]
\centering
\includegraphics[width=1\linewidth]{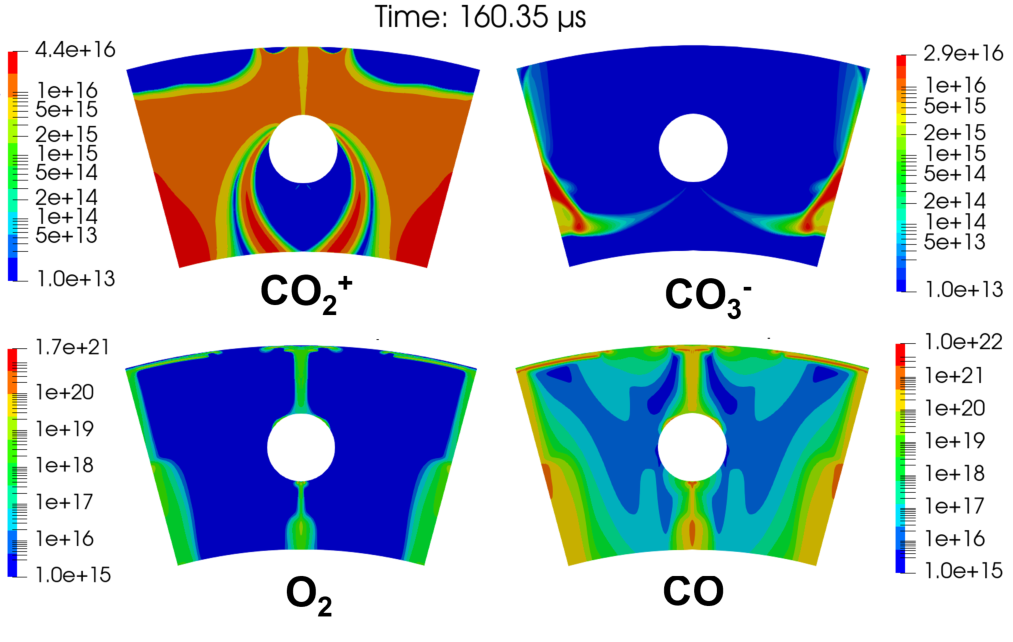} 
\caption{\label{fig:species_2ndPulse_long} \color{black} Major species density contours  (units of $\mathrm{m^{-3}}$, log-scale, min values scaled to $\mathrm{10^{13}}$ $\mathrm{m^{-3}}$ for charged species and $\mathrm{10^{15}}$ $\mathrm{m^{-3}}$ for neutral species)  at the long-term afterglow phase of the 2nd MD. O atoms (not shown here) present very similar profiles to the CO molecules.\color{black}
}
\end{figure}

In Fig.~\ref{fig:species_2ndPulse}, we plot the density contours of the dominant ion and neutral species at the decaying phase of the second current pulse. Similar to the first MD development, during this second pulse, $\mathrm{CO_{2}^{+}}$ ions are preferentially formed inside the streamer and SIW bodies with densities in the order of $\mathrm{10^{19}-10^{20}}$ $\mathrm{m^{-3}}$. $\mathrm{O^{-}}$ ions are quickly transformed to $\mathrm{CO_{3}^{-}}$ ions (R9) which again are the dominant ion species and whose density is higher inside the glow region of the MD (order of  $\mathrm{10^{18}}$ $\mathrm{m^{-3}}$). 
 The main neutral species produced by the second MD are O atoms and CO molecules, both having average densities in the order of $\mathrm{10^{20}}$ $\mathrm{m^{-3}}$ and very similar spatial distribution. The remnant neutral species from the first pulse are clearly visible in Fig.~\ref{fig:species_2ndPulse}, emphasizing the very slow neutral diffusion (and generally loss) processes compared to the AC cycle time-scales. $\mathrm{O_{2}^{-}}$ ion concentration remains negligible (lower than $\mathrm{10^{11}-10^{12}}$ $\mathrm{m^{-3}}$). In contrast, the distribution and density of $\mathrm{O_{2}}$ molecules (not shown here) remain relative unchanged as compared to the 1st pulse afterglow (Fig.~\ref{fig:species_2ndPulse_long}). As noted in the Sec.~\ref{subsubsec:1stpulse}, $\mathrm{O_{2}}$ molecules are preferentially produced inside the afterglow phase of each MD.

\begin{figure}[h]
\centering
\includegraphics[width=1\linewidth]{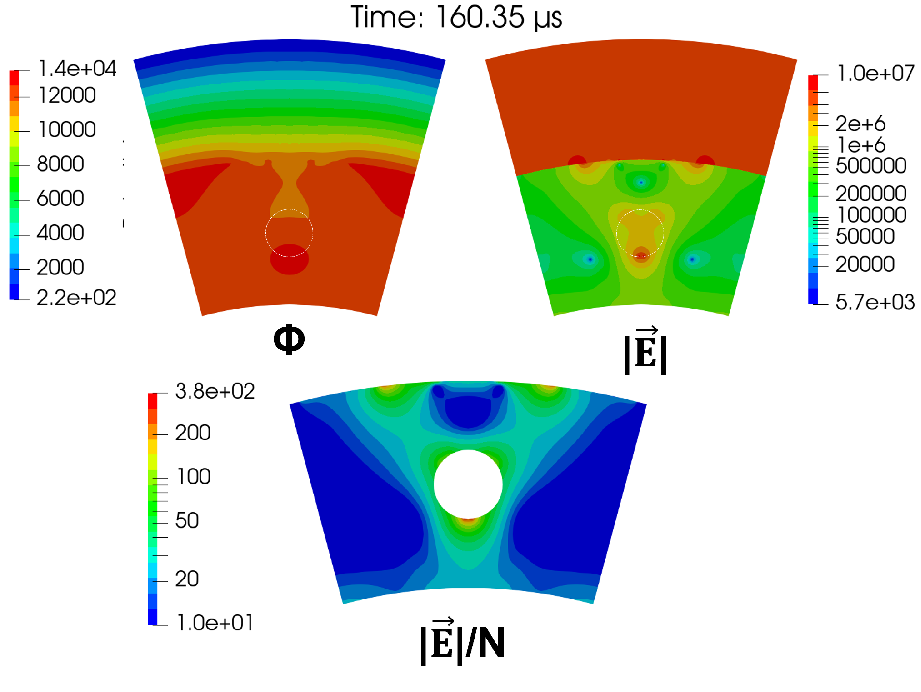} 
\caption{\label{fig:E_2ndPulse_long} \color{black} Potential $\Phi$ [V], electric field magnitude $|\vec{E}|$ [V/m] (log-scale), reduced electric field $|\vec{E}/N|$ [Td] (log-scale) and electron density, $n_{e}$, [$\mathrm{m^{-3}}$] (log-scale, min values scaled to $\mathrm{10^{13}}$ $\mathrm{m^{-3}}$ ) contours, at the long-term afterglow phase of the 2nd MD. The position of the dielectric rod is indicated with the white dashed circle.\color{black}
}
\end{figure}

During the long-term afterglow phase (Fig.~\ref{fig:species_2ndPulse_long}), positive $\mathrm{CO_{2}^{+}}$ ions fill up a large area of the reactor while charging the dielectric layer at continiously further positions, eventually merging with the remnant positively charged area from the 1st MD. The whole dielectric layer is now covered with non-uniform positive surface charges (larger values at the SIW propagation portion) which strongly modify the potential distribution in the reactor and dielectric layer. As seen in Fig.~\ref{fig:E_2ndPulse_long}, the reduced electric field obtains values around and higher than 50 Td inside a Y-shaped zone, while maximum values of around 380 Td are found mainly at the rods inner-poles.
%In later time-instants and despite the continously rising, positive, applied voltage, electron diffusion and recombination losses dominate, so that the electron density is gradually diminished on a $\mathrm{\mu}$s time-scale.  We define here, the long-term afterglow time-instant, as the instant after a current pulse when the electron density has dropped below $\mathrm{10^{12}}$ $\mathrm{m^{-3}}$ so that electrons are effectively evacuated from the reactor. As so, the long-term afterglow of major ion and neutral species at  t$=$70.849 $\mathrm{\mu}$s (approx. 22 $\mathrm{\mu}$s after the current pulse) is plotted in Fig.~\ref{fig:species_1stPulse_long}. At the same time-instant, in Fig.\ref{fig:E_1stPulse_long}, we plot the electric field magnitude, reduced electric field and potential distribution. 

 In the next subsection, we focus on the third current pulse (the first negative MD) and the discharge/species spatiotemporal development.
 
\subsubsection{Third current pulse and its afterglow}\label{subsubsec:3rdpulse}
\begin{figure}[h]
\centering
\includegraphics[width=1\linewidth]{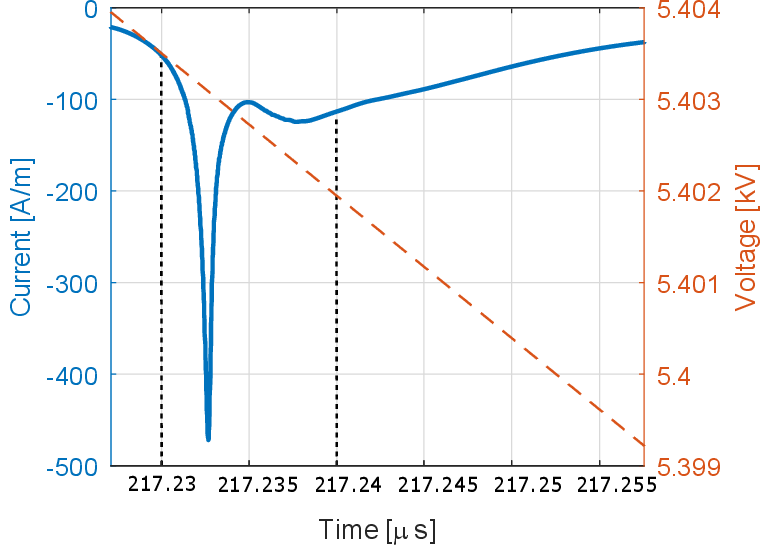} 
\caption{\label{fig:IV_3rdPulse} \color{black}Current and voltage during the 3rd MD. \color{black}
}
\end{figure}

Fig.~\ref{fig:IV_3rdPulse} depicts the I-V curve during the 3rd MD, which mainly consists of one negative current peak. Note that the applied voltage is still positive at this time-instant (around 5.4 kV). The positively charged dielectric layer causes the reversal of the potential difference between the anode (inner electrode) and the dielectric.  For the sake of brevity and length constraints of this paper, we discuss more consicely this current pulse.

\begin{figure}[h]
\centering
\includegraphics[width=1\linewidth]{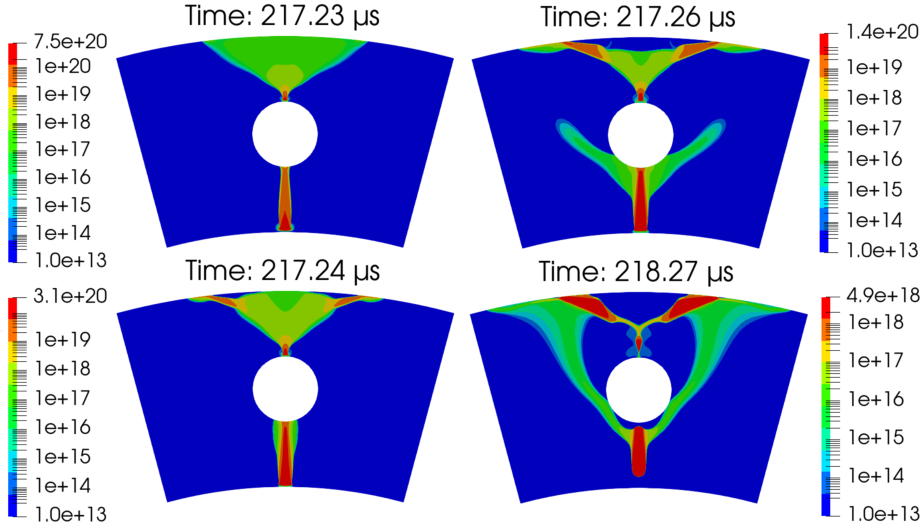} 
\caption{\label{fig:ne_3rdPulse} \color{black} Electron density contours (units of $\mathrm{m^{-3}}$, log-scale, min values scaled to $\mathrm{10^{13}}$ $\mathrm{m^{-3}}$) at different time-instants (see vertical dashed lines at Fig.~\ref{fig:IV_3rdPulse}), during and after the 3rd MD.  \color{black}
}
\end{figure}

 Fig.~\ref{fig:ne_3rdPulse} shows the electron density contours at different time-instants, during the MD development and at the current decaying phase. Initially, the MD develops in less than 5 ns as a negative streamer discharge at the inner electrode-rod region and a glow discharge at the rod-dielectric layer region. The latter glow discharge, is quite diffused and attaches to the dielectric layer portion where positive surface charge is less intense, following the reduced field distribution. It quickly transits to secondary streamers propagating towards the intense positively charged portions of the dielectric layer (remnant of SIW from previous pulse). The MD presents clearly a complex spatial profile due to the strong non-uniformities of surface charges.

\begin{figure}[h]
\centering
\includegraphics[width=1\linewidth]{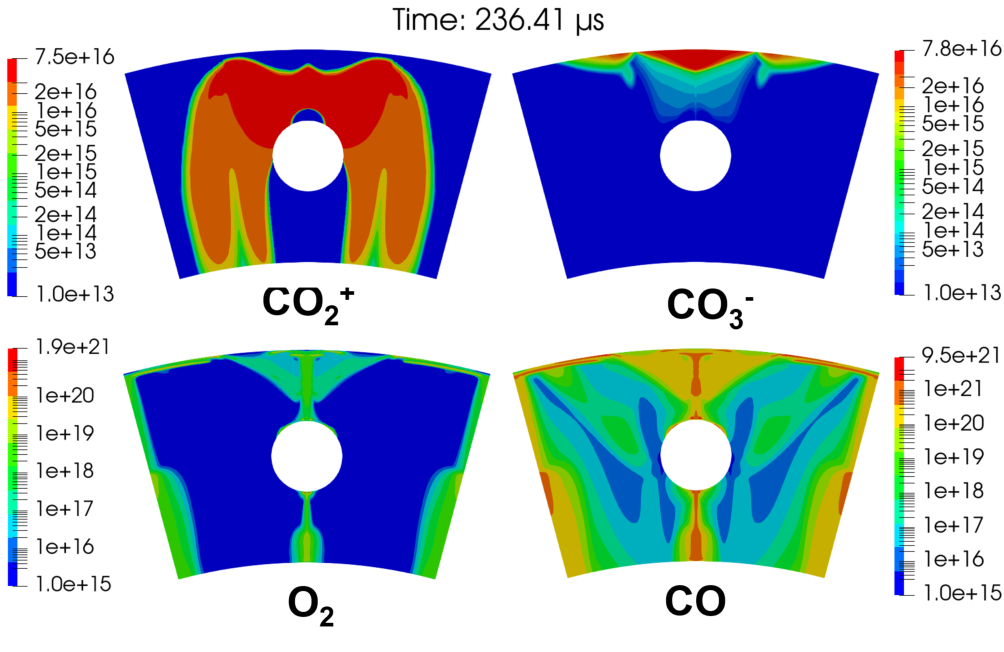} 
\caption{\label{fig:species_3rdPulse_long} \color{black} Major species density contours  (units of $\mathrm{m^{-3}}$, log-scale, min values scaled to $\mathrm{10^{13}}$ $\mathrm{m^{-3}}$ for charged species and $\mathrm{10^{15}}$ $\mathrm{m^{-3}}$ for neutral species)  at the long-term afterglow phase of the 3rd MD. O atoms (not shown here) present very similar profiles to the CO molecules.\color{black}
}
\end{figure}

\begin{figure}[h]
\centering
\includegraphics[width=1\linewidth]{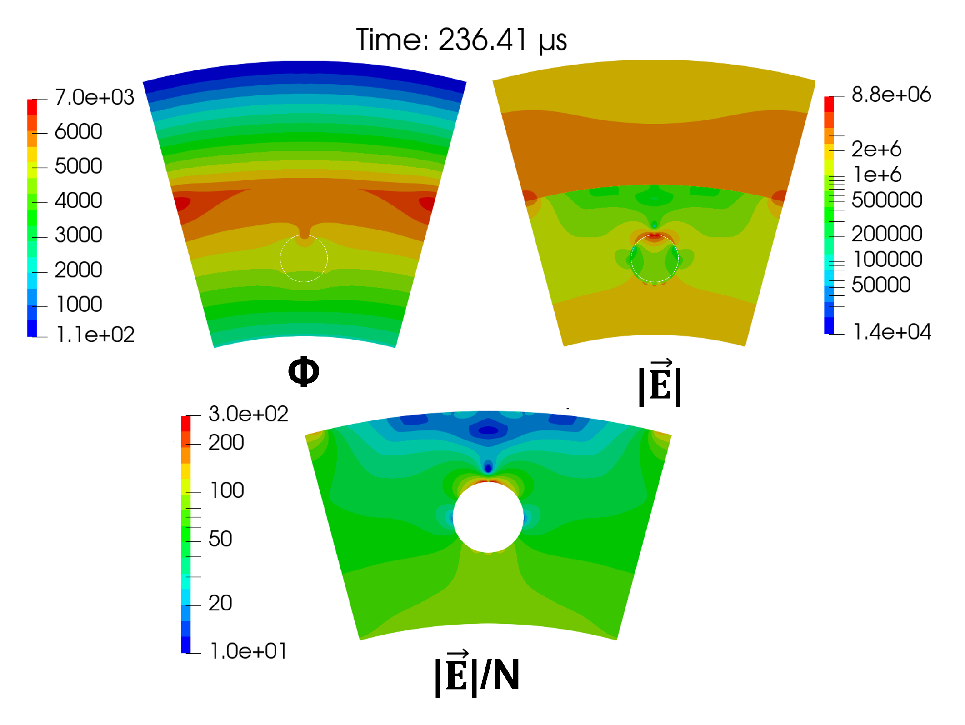} 
\caption{\label{fig:E_3rdPulse_long} \color{black} Potential $\Phi$ [V], electric field magnitude $|\vec{E}|$ [V/m] (log-scale), reduced electric field $|\vec{E}/N|$ [Td] (log-scale) and electron density, $n_{e}$, [$\mathrm{m^{-3}}$] (log-scale, min values scaled to $\mathrm{10^{13}}$ $\mathrm{m^{-3}}$ ) contours, at the long-term afterglow phase of the 3rd MD. The position of the dielectric rod is indicated with the white dashed circle.\color{black}
}
\end{figure}

The long-term afterglow of major ion and neutral species at t$=$70.849 $\mathrm{\mu}$s (approx. 22 $\mathrm{\mu}$s after the current pulse) is plotted in Fig.~\ref{fig:species_3rdPulse_long}. Positive $\mathrm{CO_{2}^{+}}$ ions drift towards the inner electrode and their population is very low inside the inner electrode-rod regions. Negative $\mathrm{CO_{3}^{-}}$ ions follow the opposite direction, drifting towards the less charged portion of the dielectric layer (inside the rod-dielectric region). The distributions of $\mathrm{CO}$ and $\mathrm{O_{2}}$ species, follow the MD developments in a similar way to the previous MDs. 
At the same time-instant, in Fig.\ref{fig:E_3rdPulse_long}, we plot the electric field magnitude, reduced electric field and potential distribution. The reduced electric field is lower than 100 Td inside the reactor except a thin zone of high E/N (> 200 Td) near the outer pole of the dielectric rods and a zone of low E/N (< 20 Td) at the rod-dielectric regions and close to the dielectric layer. The potential distribution indicates that negative charges are now dominating the surface charging process of the dielectric layer especially closer to the rod positions, reducing the positively charged layers density. 

It is clear that each MD presents similar glow-streamer-SIW transitions but their spatiotemporal behavior and intensity is heavily governed by the non-uniform space and (mainly) surface charge distributions, which give rise to non-repeatable breakdown phases, MD development and afterglow phases.  In the next subsection, and again for the sake of brevity, we discuss the 4th, 5th and 6th MD, putting emphasis mainly on the afterglow phase of each pulse. The reader can find a video file of the detailed electron density and CO evolution inside the first AC cycle, in the supplementary material.  

\subsubsection{Subsequent current pulses and their afterglow characteristics}\label{subsubsec:restofpulses}
\begin{figure}[h]
\centering
\includegraphics[width=1\linewidth]{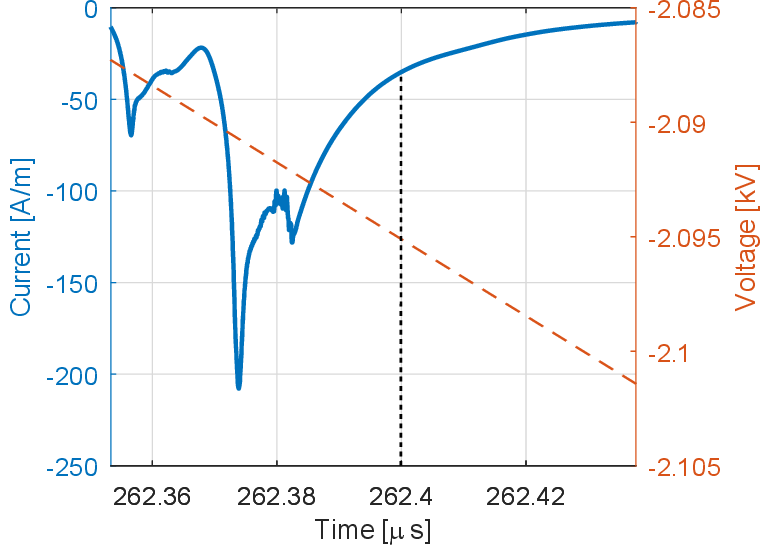} 
\caption{\label{fig:IV_4thPulse} \color{black}Current and voltage during the 4th MD. \color{black}
}
\end{figure}

Fig.~\ref{fig:IV_4thPulse} depicts the I-V curve during the 4th MD, which consists of three distinct current peaks.

\begin{figure}[h]
\centering
\includegraphics[width=1\linewidth]{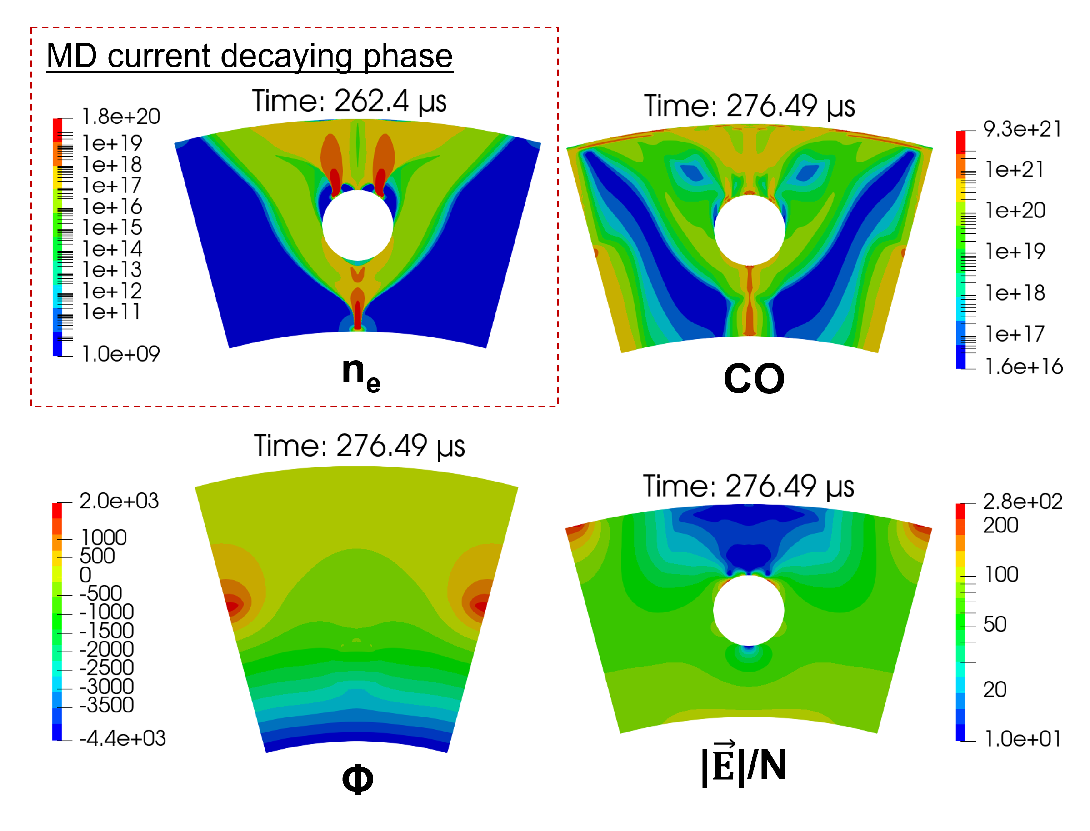} 
\caption{\label{fig:4thPulse_brief} \color{black} Top-left: Electron density contours  (units of $\mathrm{m^{-3}}$, log-scale, min values scaled to $\mathrm{10^{13}}$ $\mathrm{m^{-3}}$) at the decaying phase of the 4th MD current pulse. Rest:  CO density contours  (units of $\mathrm{m^{-3}}$, log-scale, min values scaled to $\mathrm{10^{15}}$ $\mathrm{m^{-3}}$),  potential $\Phi$ [V] and reduced electric field $|\vec{E}/N|$ [Td] (log-scale) at the long-term afterglow phase of the 4th MD.\color{black}
}
\end{figure}

In Fig.~\ref{fig:4thPulse_brief}, we plot the electron density contours (top-left) at the current decaying phase. Three distinct negative streamer discharges are visible. The first occurs at the inner electrode-rod regions while the rest occur at the rod-dielectric regions but in contrast to the previous pulse, the attachement to the dielectric rods present a symmetric offset form its outer pole position. In the same figure, we plot the afterglow contours of CO number density, potential and reduced electric field. Once again, the MD distribution reflects on the CO production zones while the dielectric layer now posseses positive surface charges only at a small (discontinuous) portion of its total arc-length (regions where the potential is largely positive). The reduced electric field is higher at these regions (> 200 Td), while it remains lower than 100 Td inside the rest of the reactor and lower than 20 Td in the rod-dielectric regions.

\begin{figure}[h]
\centering
\includegraphics[width=1\linewidth]{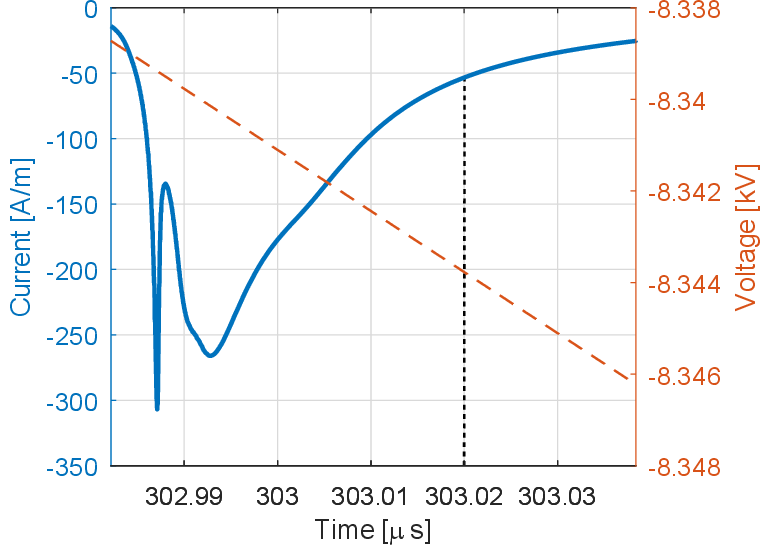} 
\caption{\label{fig:IV_5thPulse} \color{black}Current and voltage during the 5th MD. \color{black}
}
\end{figure}

Fig.~\ref{fig:IV_5thPulse} depicts the I-V curve during the 5th MD, which consists of three distinct current peaks.

\begin{figure}[h]
\centering
\includegraphics[width=1\linewidth]{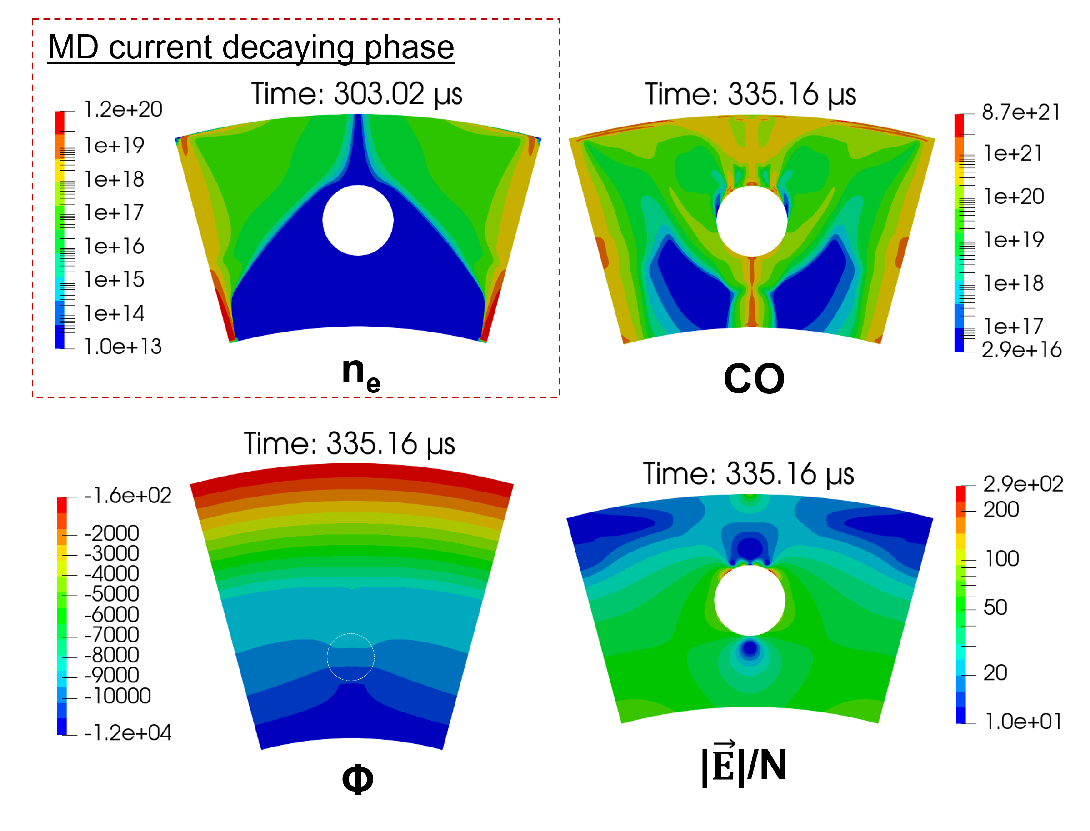} 
\caption{\label{fig:5thPulse_brief} \color{black} Top-left: Electron density contours  (units of $\mathrm{m^{-3}}$, log-scale, min values scaled to $\mathrm{10^{13}}$ $\mathrm{m^{-3}}$) at the decaying phase of the 5th MD current pulse. Rest:  CO density contours  (units of $\mathrm{m^{-3}}$, log-scale, min values scaled to $\mathrm{10^{15}}$ $\mathrm{m^{-3}}$),  potential $\Phi$ [V] and reduced electric field $|\vec{E}/N|$ [Td] (log-scale) at the long-term afterglow phase of the 5th MD.\color{black}
}
\end{figure}

As seen in Fig.~\ref{fig:5thPulse_brief}, at the current decaying phase (top-left), the discharge is similar (but opposite in polarity) with the 2nd MD, occuring at the azimuthal region between the dielectric rods: a dense (approx. $\mathrm{10^{20}}$ $\mathrm{m^{-3}}$) streamer is formed near the inner electrode. In contrast with the positive-carrying current discharge of the 2nd MD which quickly transits to a SIW at the dielectric layer, the negative surface discharge is weaker (approx. $\mathrm{10^{18}}$ $\mathrm{m^{-3}}$) and it mostly propagates as an electron cloud, contributing negatively to the surface dielectric charge. This effect is obvious in the long-term afterglow potential and reduced electric field contours (same figure): while the applied potential at the inner electrode is strongly negative at this time instant (-1.2 kV), E/N is lower than 70 Td in most of the reactor volume (as surface charging homogenizes the electric potential therein), while presenting zones of very low E/N near the surface discharges and the inner and outer poles of the rods. CO is again produced mainly inside the MD zones, similar to previous MD pulses.

\begin{figure}[h]
\centering
\includegraphics[width=1\linewidth]{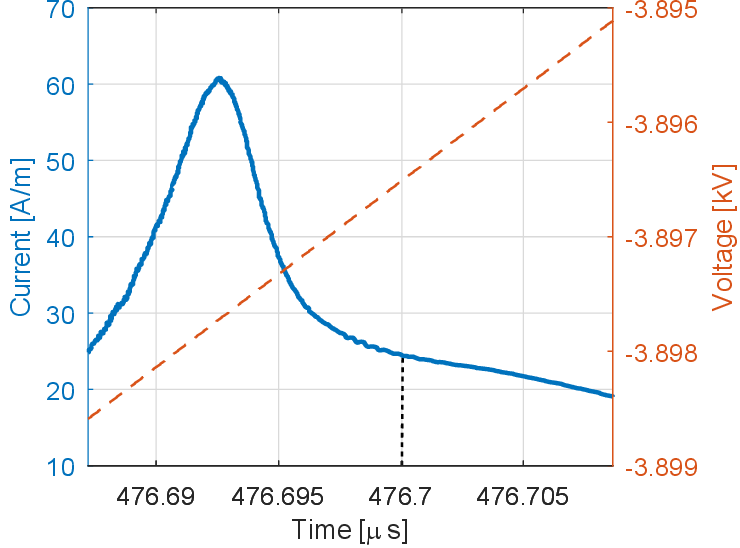} 
\caption{\label{fig:IV_6thPulse} \color{black}Current and voltage during the 6th MD. \color{black}
}
\end{figure}

Lastly, Fig.~\ref{fig:IV_6thPulse} depicts the I-V curve during the 6th MD, which consists of one smooth current peak. We note that this current pulse is positive despite the negative values of the applied voltage (approx. -3.89 kV) which indicates another reversal of the potential difference inside the reactor due to memory effects. 

\begin{figure}[h]
\centering
\includegraphics[width=1\linewidth]{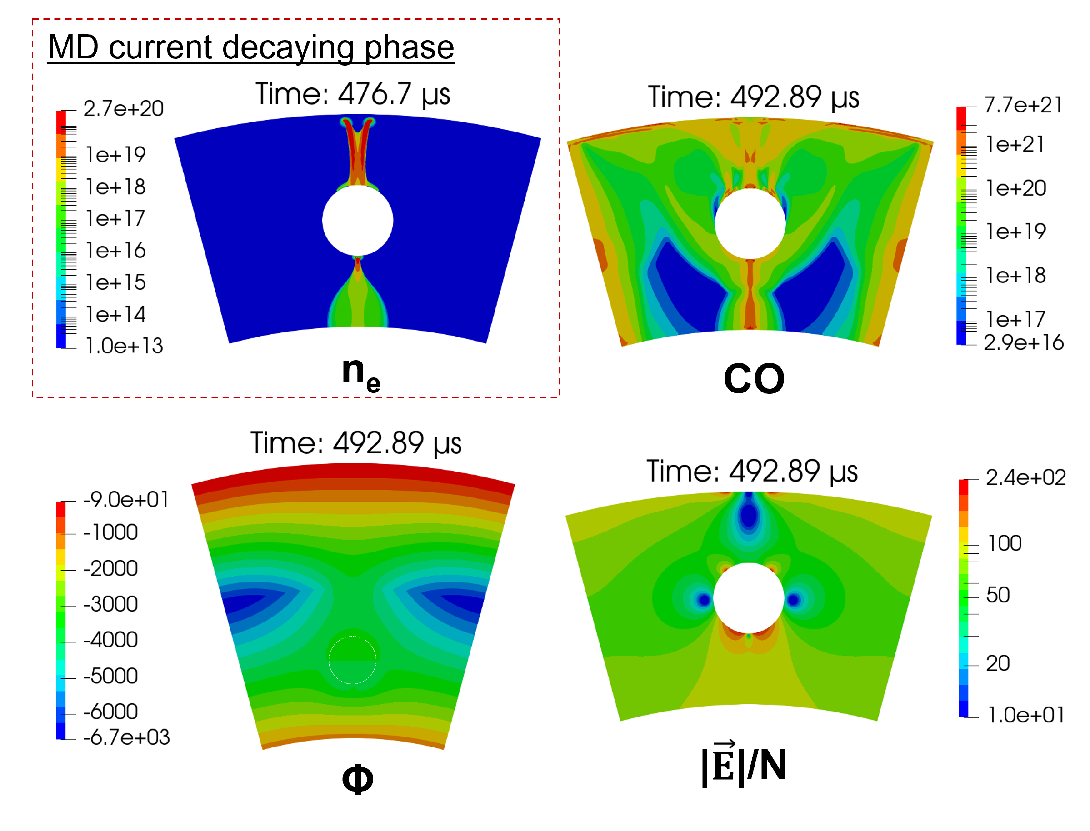} 
\caption{\label{fig:6thPulse_brief} \color{black} Top-left: Electron density contours  (units of $\mathrm{m^{-3}}$, log-scale, min values scaled to $\mathrm{10^{13}}$ $\mathrm{m^{-3}}$) at the decaying phase of the 6th MD current pulse. Rest:  CO density contours  (units of $\mathrm{m^{-3}}$, log-scale, min values scaled to $\mathrm{10^{15}}$ $\mathrm{m^{-3}}$),  potential $\Phi$ [V] and reduced electric field $|\vec{E}/N|$ [Td] (log-scale) at the long-term afterglow phase of the 6th MD.\color{black}
}
\end{figure}

As seen in Fig.~\ref{fig:6thPulse_brief}, at the current decaying phase (top-left), the last MD of the first AC cycle is quite similar with the 1st MD in terms of breakdown position. It consists of a glow-like discharge in the inner electrode-rod regions and, inside the rod-dielectric region, a streamer discharge which branches into a forked tongue structure as it approaches the dielectric layer.   Similar to the 1st MD, a rather short propagating SIW at the dielectric layer is observed. 
Through the mechanism of positive ion drifting and jumping along the dielectric layer, the latter is charged positively as it is clearly seen in the long-term afterglow potential and reduced electric field contours (same figure): the restructuring of the potential distribution leads to complex partners of low/high E/N at the perimeter of the rods.

We note here that memory effects (due to both dielectric layer and rods charging), disturb the potential distribution inside the reactor in such way as to reinitiate the cyclic breakdown process in earlier time (occurence of positive MD before the end of the AC cycle) . To this respect, a quasi-periodic regime is expected to be established in several AC periods (2-3 periods normally~\cite{kourtzanidis2020self}). In the next section, we report on cycle-averaged and spatial average quantities.

%FROM Zhang, Shen, et al. "Numerical investigation on the effects of dielectric barrier on a nanosecond pulsed surface dielectric barrier discharge." Molecules 24.21 (2019): 3933.
%***The effects of epsilon on the distributions of wall charge density at 15 ns are consequently studied and shown in Figure 11. Because of the increasing E, the densities of electrons and ions increase with increasing epsilon, and the charge accumulation on the dielectric surface is enhanced. In addition, the dielectric is stronger polarized due to higher epsilon, more charge is induced inside and on the surface of the dielectric barrier. However, the charge on the surface to the exposed electrode is nearly the same.***

\subsection{Cycle-averaged species distribution}\label{subsec:average}
\begin{figure}[h]
\centering
\includegraphics[width=1\linewidth]{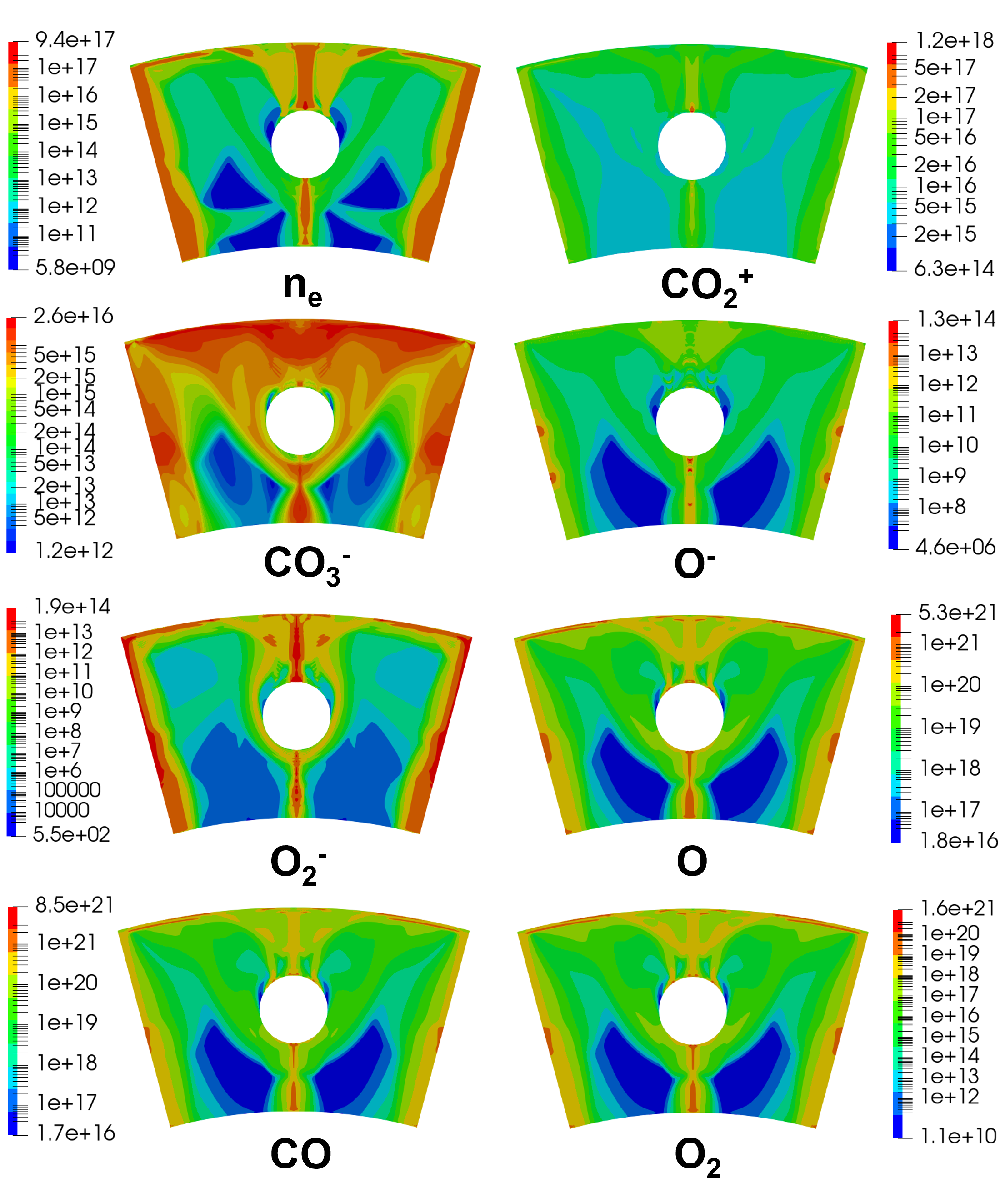} 
\caption{\label{fig:av_12rods} \color{black} Species cycle-average number density contours (in $\mathrm{m^{-3}}$ and log scale). \color{black}
}
\end{figure}

%In order to further elucidate the various discharge regimes and its great spatiotemporal variance inside an AC cycle, we have performed simulations with a coarser rod packing density. Twelve (12) rods are used in this case-study. 
Fig.~\ref{fig:av_12rods} depicts the cycle-average number density contours of all species included in the current model (except of $\mathrm{CO_{2}}$ species, whose density is almost unchanged in the 1st AC period). All species follow more or less the MD developments in terms of spatial distribution. $\mathrm{CO_{2}^{+}}$ ions present a quite homogeneous average distribution, with densities in the $\mathrm{10^{15}-10^{16}}$ $\mathrm{m^{-3}}$ range. Consistent with the detailed spatiotemporal analysis of the previous sections, $\mathrm{CO_{3}^{-}}$ is found to be the dominant negative ion whose density exceeds by two orders of magnitude the $\mathrm{O_{2}^{-}}$ and $\mathrm{O^{-}}$ density. The distributions of neutral species are very similar but it is worth mentioning that CO and O species exist in densities higher than $\mathrm{10^{19}}$ $\mathrm{m^{-3}}$ in nost of the reactor volume while  $\mathrm{O_{2}}$ are in the $\mathrm{10^{15}-10^{16}}$ $\mathrm{m^{-3}}$ range.

\begin{figure}[h]
\centering
\includegraphics[width=1\linewidth]{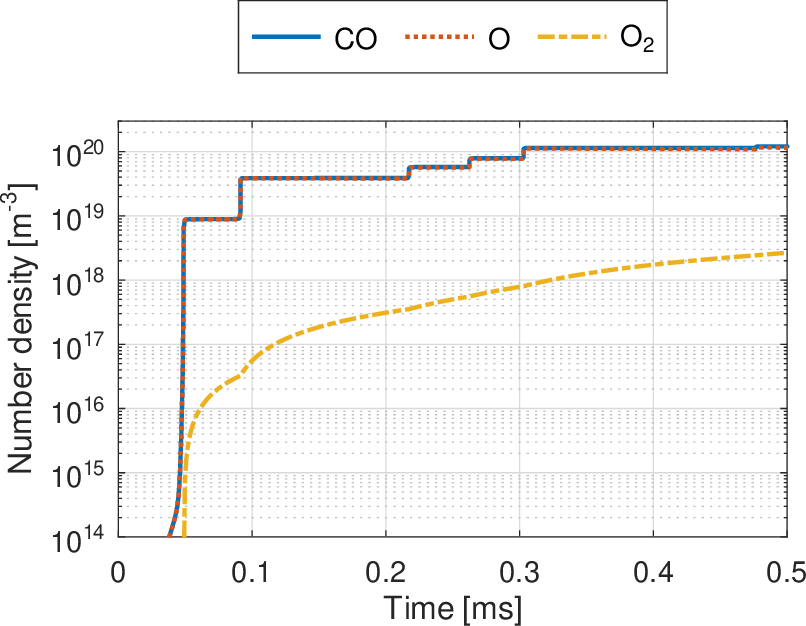} 
\caption{\label{fig:neutralspecies_vs_time} \color{black} Spatial average (mean) number density ($\mathrm{m^{-3}}$) for all neutral species vs time.  \color{black}
}
\end{figure}

To further elucidate the species production, in Fig.~\ref{fig:neutralspecies_vs_time}, the spatial average (mean) number density of all neutral species is plotted versus time. The results confirm that CO and O species are produced during each MD development owing to dense streamer discharges (note the density jump at each current pulse time-instant) while $\mathrm{O_{2}}$ density presents a smooth increase which relates their production with the glow, current decaying and afterglow phases of each MD. 

\begin{figure}[h]
\centering
\includegraphics[width=1\linewidth]{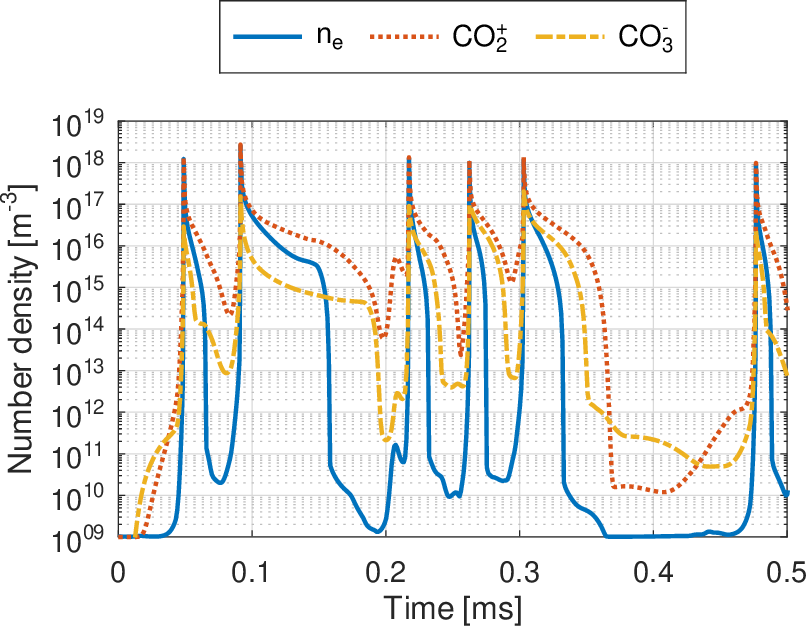} 
\caption{\label{fig:chargedspecies_vs_time} \color{black} Spatial average (mean) number density  ($\mathrm{m^{-3}}$)  for dominant charged species vs time.  \color{black}
}
\end{figure}

\begin{figure}[h]
\centering
\includegraphics[width=1\linewidth]{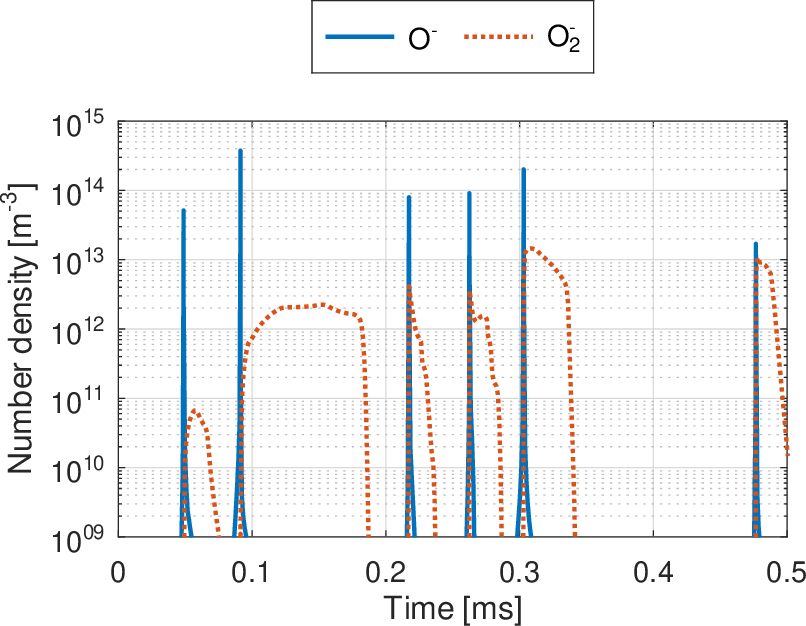} 
\caption{\label{fig:chargedspecies_2_vs_time} \color{black} Spatial average (mean) number density  ($\mathrm{m^{-3}}$)  for rest of charged species vs time.  \color{black}
}
\end{figure}

In Fig.~\ref{fig:chargedspecies_vs_time} and Fig~\ref{fig:chargedspecies_2_vs_time},  the spatial average (mean) number density of all charged species is plotted versus time. The dominant charged species (Fig.~\ref{fig:chargedspecies_vs_time}) are mainly produced during each MD development and slowly relax towards each afterglow phase with a relaxation time-scale in the order of 10-100 $\mu$s. Our calculations show that there is no strong electron accumulation inside the reactor even during the subsequent and closely spaced in time, negative MDs. Nevertheless, during the current-decaying and afterglow phases, a persisting net positive volumetric charge exists, with a longer associated relaxation time during MDs which ignite long SIWs (2nd and 5th MD). 
As we've already seen, $\mathrm{O^{-}}$ ions produced during each MD (mainly via R3) are quickly transformed to $\mathrm{CO_{3}^{-}}$ ions (via R9). On the other hand, $\mathrm{O_{2}^{-}}$ ions are mainly produced during the current decaying and afterglow phases of each MD (via R7) following the growing population of $\mathrm{O_{2}}$ molecules. Their densities though remain relatively low inside the whole AC cyle, despite the growing and accumulating tendency of $\mathrm{O_{2}}$ molecules. The temporal behaviour of $\mathrm{O_{2}^{-}}$, denotes that inside the first AC period, R7 is not an efficient reaction and consequently R11 does not contribute efficiently to the neutral (and ions) species balance. This claim needs revisiting as $\mathrm{O_{2}}$ densities are expected to increase in further AC cycles (until they reach a quasi-steady state), enhancing the $\mathrm{O_{2}^{-}}$ production via R7.

\begin{figure}[h]
\centering
\includegraphics[width=1\linewidth]{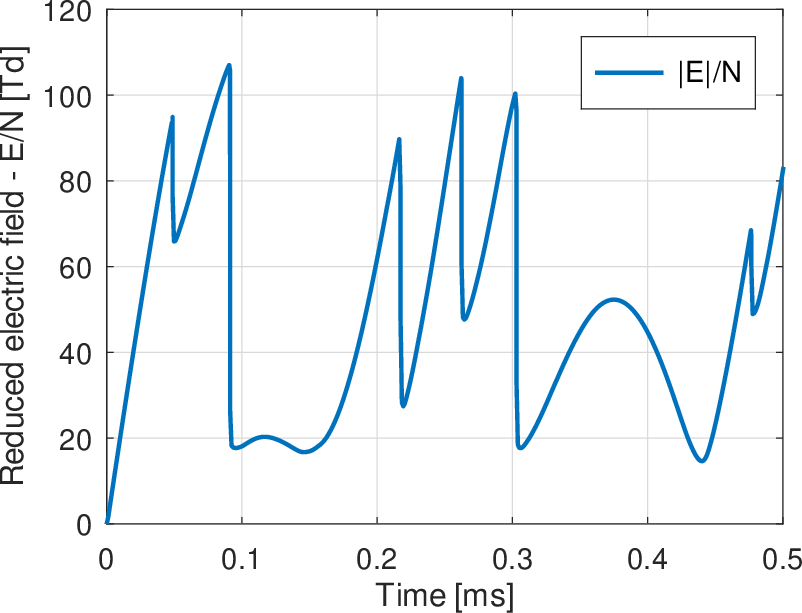} 
\caption{\label{fig:electric_vs_time} \color{black} Spatial average (mean) reduced electric field ($|\vec{E}|/N$ in Td units) in the reactor vs time.  \color{black}
}
\end{figure}

Last, in Fig.~\ref{fig:electric_vs_time}, we plot the spatial average (mean) reduced electric field in the reactor volume versus time. Mean values of E/N range from approx. 18 - 108 Td. Each current pulse enhances the mean electric field in the reactor, which subsequently relaxes to lower values due to surface charging and potential redistribution.

\section{Discussion}\label{sec:discussion}
We summarize below and elaborate on the main assumptions and limitations of the model used in this study. In addition, we propose several future directions mainly based on the numerical efforts and advancements required for a complete and accurate picture of plasma-based $\mathrm{CO_{2}}$ conversion. 
\begin{enumerate}
\item{Photoionization and background ionization:} A very recent work used a 2D-PIC model to investigate the effects of photoionization on positive CO2 streamers~\cite{li2023effect}. The authors demonstrate that photoionization plays a role in sustaining a positive streamer but they've also confirmed the relatively low effects of photoionization compared to air discharges and the requirement of high backgroung electric fields (close to the breakdown threshold) for sustaining the discharge through this mechanism. In addition, they comment on the large uncertainties in the available parameters required for any numerical model to describe the process in CO2 (absorption length, photon production source, etc). To this end, the exclusion of photoionization in our model should mainly influence the positive MD and SIW initiation and propagation velocities but the results obtained should remain qualitatively similar. The (low) background ionization levels used in the simulations are sufficient for breakdown and sustaining the different discharge regimes under the conditions encountered in this work. More detailed studies (based on accurate measurements of photoionization parameters in atmospheric conditions) are definitely needed towards a better understanding of such phenomena.
\item{Plasma chemistry and gas temperature:} As noted in Sec.~\ref{sec:chemistry}, the plasma chemistry model is quite simplified. The vibrational kinetics are not explicitly included in the model and thus the ladder-climbing process for dissociation is excluded and the dissociation degree is expected to be underestimated to a limited extent~\cite{kozak2014splitting}. Moreover, various neutral, ion-neutral and  $\mathrm{O_{2}}$ related electron impact reactions as well as $\mathrm{O_{3}}$ formation and related reactions are not part of our chemistry model. Despite these facts, the reduced model has been proven adequate for simulating atmospheric pressure $\mathrm{CO_{2}}$ DBDs and the main results of this work (multi-regime plasma spatiotemporal development) should not be largely affected. We propose a systematic description of all kinetic and transport rate data (sources, conditions used for solving the two-term Boltzmann equation, etc) by the various active reserach groups in the field in order to minimize uncertainties, foster comparability of the obtained numerical results and promote simple validation studies with experiments. Lastly, the effects of gas temperature on reaction rates and eventual discharge periodic regime deserves a separate study. We note that measurements in AC-DBDs show that the gas temperature typically lies near the ambient values and as such, we expect this influence to be limited. 
%\item{SEE and other surface kinetics:} Surface kinetics are only included here as direct recombination of all charged species in dielectric and metal surfaces. The typical formulation for including the SEE kinetics through the positive ion boundary flux and SEE coefficient is subject to uncertainties in both the ion energy as well the values of the coefficient itself (which depends on the surface material and structure). A more elaborated parametric study on the influence of the SEE coefficient in the discharge development (and mainly it sustainability) should be performed under the specific conditions studied herein. Such study should take into account the catalyst layers if possible and as such, we suggest a collaborative effort of research groups in the interaction between material science, plasma physics and chemical engineering. 
\item{Three dimensional (3D) effects} A coaxial PB-DBD reactor operates clearly in a 3D mode. By neglecting the spatial evolution of the discharges in the longitudinal direction of the reactor (across its length), we assume a homogenous, unifom discharge along this dimension. This assumption inherent to 2D models, not only does not allow for a proper description of the total number of discharge filaments inside the whole reactor but also affects the 2D discharge behavior and the estimation of the conversion rate. The discharge filaments in the third dimension, will experience annihilation and possibly self-organization based on electrostatic repulsion, 3D surface charging effects as well as travelling behaviour~\cite{boeuf2012generation}, leading to a non-uniform distribution. The overall dynamics of a single three-dimensional filament can be reflected only to a certain extent with a 2D approximation. Moreover, as we've seen, the filamentary discharge plays a crucial role on $\mathrm{CO_{2}}$ conversion and thus these three-dimensional spatiotemporal effects should not be neglected. In addition, in a typical reactor the gas flows in the longitudinal direction and thus the residence time (or better the time duration that each molecule experiences each filament and is thus subject to plasma related reactions) is directly linked to this spatial distribution and the filament thickness in the third dimension. We argue that the deep comprehension of such effects in simplified geometries where the discharge filaments can be controlled and measured accurately, should be an important focus of both experimental and numerical research groups in order to optimize the conversion process. Acknowledging the extremely demanding nature of 3D plasma simulations, we suggest once again a collaborative effort towards this target where precise cross-validation with simple experiments can lead to accurate simplifications and extensions of 2D (or even 1D/0D) models without the extensive need of tuning parameters.
%\item{Memory effects and residual charge:} Memory effects due to the dielectric charging but also due to residual charges from one AC cycle to the next will modify the discharge behavior until a quasi-steady state is established. Preliminary calculations with an oversimplified chemistry demonstrate that during the 2nd and 3rd AC periods the positive MDs increase to a total of four, while only two negative MDs are present per AC period.  In addition, due to the redistribution of charged but also neutral species densities, the MDs intensity vary. We intent to capture the quasi-steady regime by allowing the calculation to continue for multiple AC periods and report the results in a future publication.  
\end{enumerate}

\section{Conclusion and future directions}\label{sec:conclusion}
A two-dimensional self-consistent model of AC-DBD excited pure CO2 plasma has been developed and used to simulate the spatiotemporal characteristics of the discharge inside the first AC cycle. The numerical results indicate that the plasma operates in glow, streamer and SIW modes which present significant 
spatial inhomogeneities owing to the surface charging and residual space charge effects. The dominant negative ions in the discharge are $\mathrm{CO_{3}^{-}}$, while the population of $\mathrm{O^{-}}$ and $\mathrm{O_{2}^{-}}$ ions are more than two orders of magnitude lower. Volumetric and surface streamers (SIW) are found to be the most efficient discharge modes for direct $\mathrm{CO_{2}}$ dissociation to O atoms and CO molecules. $\mathrm{O_{2}}$ molecules are mainly produced after each MD development. The numerical results indicate that an increase in the microdischarge number and intensity would be beneficial for overall $\mathrm{CO_{2}}$ conversion efficiency, reducing the $\mathrm{O_{2}}$ production phases (and thus reducing the backward (recombination) reaction of CO with O2 molecules towards  $\mathrm{CO_{2}}$).
A detailed study on packing density, dielectric/electrode material properties and frequency/voltage dependence is necessary in the future in order to parametrize the PB-DBD reactor. In addition, extension of the simulations into multiple AC cycles taking into account memory effects could reveal the quasi-steady state of the reactor operation and its deviation from the 1st AC cycle dynamics, also identifying key timescales and more accurate cycle-averaged quantities that could directly feed global models. Lastly, the influence of gas temperature is left for a future study.

\section{Acknowledgments}
The author would like to thank F. Rogier, G. Dufour and T.D. Nguyen (ONERA - Toulouse, France) for the collaborative effort in the continous development of the COPAIER plasma solver.

\section*{References}

\bibliography{PDBD_Kourtzanidis_CLEAN}%{}
\bibliographystyle{unsrt}

\end{document}